\documentclass[a4paper,12pt,english]{article}

\usepackage{amsfonts,bm,amssymb,euscript,array,babel,cite}
\usepackage{mathtools}
\usepackage{tikz-cd}
\usepackage{amsmath}
\usepackage{hhline}
\usepackage[amsmath]{empheq}
\usepackage{hyperref}
\usepackage{cancel}
\usepackage{mathrsfs}
\usepackage{pdfsync}
\usepackage{tikz}

\newsavebox{\mysaveboxM}
\newsavebox{\mysaveboxT}

\makeatletter

\newcommand{\dd}{\mathrm{d}}

\newcommand{\w}{\wedge}
\newcommand{\cbral}{[\![}
\newcommand{\cbrar}{]\!]}

\newcommand{\be}{\begin{equation}}
\newcommand{\ee}{\end{equation}}
\newcommand{\sfrac}[2]{{\textstyle\frac{#1}{#2}}}
\def\nn{\nonumber}
\def \bea{\begin{eqnarray}} 
\def\eea{\end{eqnarray}}
\def\bse{\begin{subequations}}	
\def\ese{\end{subequations}}
\def\bal{\begin{align}} 
\def\eal{\end{align}}

\def\bi{\begin{itemize}} 
\def\ei{\end{itemize}}

\makeatother

\def\a{\alpha} \def\b{\beta} \def\g{\gamma} \def\G{\Gamma} \def\d{\delta}

\def\s{\sigma} \def\S{\Sigma}

\def\R{{\mathbb R}}

\def\one{\mbox{1 \kern-.59em {\rm l}}}

\numberwithin{equation}{section}

\begin{document}

\makeatother
\parindent=0cm
\renewcommand{\title}[1]{\vspace{10mm}\noindent{\Large{\bf #1}}\vspace{8mm}} \newcommand{\authors}[1]{\noindent{\large #1}\vspace{5mm}} \newcommand{\address}[1]{{\itshape #1\vspace{2mm}}}

\begin{titlepage}

\begin{flushright}
 {\small\sf
 LMU--ASC 06/19\\
 MPP--2019--6\\
 EMPG--19--01}
\end{flushright}

\begin{center}

\title{ {\Large {Fluxes in Exceptional Field Theory  \\\vspace{6pt} and Threebrane Sigma-Models}}}

 \authors{\large Athanasios {Chatzistavrakidis}{$^{\a,\b,}$}{\footnote{Athanasios.Chatzistavrakidis@irb.hr}} \ \ \ \
Larisa Jonke$^{\a,\b,}${\footnote{larisa@irb.hr}} \\ \vspace{6pt} Dieter
L\"ust$^{\b,\g,}${\footnote{dieter.luest@lmu.de }} \ \ \ \ Richard J. Szabo$^{\d,}${\footnote{R.J.Szabo@hw.ac.uk}}     }
 
 \vskip 3mm
 
  \address{ $^{\a}$ Division of Theoretical Physics, Ru\dj er Bo\v skovi\'c Institute \\ Bijeni\v cka 54, 10000 Zagreb, Croatia 
  	\\[5pt]
  	$^{\b}$ Arnold Sommerfeld Center for Theoretical Physics\\ Department f\"ur Physik, Ludwig-Maximilians-Universit\"at M\"unchen\\ Theresienstra\ss e 37, 80333 M\"unchen, Germany
\\[5pt]
  	$^{\g}$ Max-Planck-Institut f\"ur Physik, Werner-Heisenberg-Institut\\ F\"ohringer Ring 6, 80805 M\"unchen, Germany
  	\\[5pt]
  	$^{\d}$ Department of Mathematics, Heriot-Watt University \\ Colin Maclaurin Building, Riccarton, Edinburgh EH14 4AS, UK \\ \vspace{4pt} Maxwell Institute for Mathematical Sciences, Edinburgh, UK \\ \vspace{4pt} The Higgs Centre for Theoretical Physics, Edinburgh, UK
}

\vskip 1cm

\begin{abstract}
\noindent
Starting from a higher Courant bracket associated to exceptional
generalized geometry, we provide a systematic derivation of all types
of fluxes and their Bianchi identities for four-dimensional
compactifications of M-theory. We show that these fluxes may be
understood as generalized Wess-Zumino terms in certain topological threebrane
sigma-models of AKSZ-type, which relates them to the higher structure of a
Lie algebroid up to homotopy. This includes geometric
compactifications of M-theory with $G$-flux and on twisted tori, and
also its compactifications with non-geometric $Q$- and $R$-fluxes in
specific representations of the U-duality group~$SL(5)$ in exceptional
field theory. 

\end{abstract}

\end{center}

\vskip 2cm

\end{titlepage}

\setcounter{footnote}{0}

{\baselineskip=12pt
\tableofcontents
}

\bigskip

\section{Introduction}
\label{sec1}

A general string or M-theory background may carry non-vanishing vacuum
expectation values for some of its NS--NS or RR field strengths,
commonly known as background fluxes which are widely used in all
modern attempts to relate string theory to low-energy phenomenology
\cite{Grana:2005jc}. Apart from the standard geometric settings, where
one may also include the possibility of non-vanishing torsion, the T-
and U-dualities of closed string and M-brane theories reveal the
existence of exotic fluxes that cannot be described in the context of
standard geometry. These are commonly known as non-geometric fluxes;
see \cite{Plauschinn:2018wbo} for a recent review in the context of
string theory with a complete list of references, and also~\cite{Szabo:2018hhh} for
a review of some of the mathematical features in the setting of the
present paper.

In the context of M-theory, the full set of fluxes for its
seven-dimensional compactifications{\footnote{For clarity, the seven
    dimensions here refer to the external spacetime.}} was determined
in \cite{Blair:2014zba} using $SL(5)$ exceptional field theory
\cite{Berman:2010is}, and studied further also for dimensions up to
seven in
\cite{Bosque:2016fpi,Gunaydin:2016axc,Kupriyanov:2017oob,Lust:2017bgx,Lust:2017bwq}. The
latter is the M-theory analogue of double field theory
\cite{dft1,dft2,dft3}, in that both theories are proposals for a
duality-invariant formulation, be it T-duality in the string theory
case or U-duality in the M-theory case. This exceptional field theory
is related by construction to a generalized geometry on the tangent
bundle extended by 2-forms \cite{HullEGG,Pacheco}. This bundle can be
equipped with a bracket \cite{Hagiwara,BZ,Zambon,Bi,Bouwknegt}, the
higher analogue of the Courant bracket defined in \cite{courant}
whose properties are collected in the structure of a Courant algebroid
\cite{liu}. In string theory, the Courant bracket was used to
systematically determine the general expressions for the full set of
geometric and non-geometric fluxes together with their Bianchi
identities \cite{Halmagyi,Blumenhagen:2012pc}; indeed, one can subsequently show that these expressions coincide with the local form of the axioms of a Courant algebroid. 

On the other hand, the axioms of a Courant algebroid also coincide
with the conditions for gauge (or BRST) invariance and on-shell
closure of the algebra of gauge transformations for a first-order
action functional for Wess-Zumino terms in three dimensions, called
the Courant sigma-model
\cite{Ikeda:2002wh,Park:2000au,Hofman:2002jz,dee1,dee2}. This may be
neatly phrased in the language of the BV field-antifield formalism
\cite{Henneaux:1989jq,Gomis:1994he}, where the structure of a general
gauge theory is encoded in the master equation: The axioms of a
Courant algebroid are equivalent to the classical master equation of
the Courant sigma-model. The Courant sigma-model falls in the general
class of topological sigma-models constructed geometrically in
\cite{Alexandrov:1995kv}, called AKSZ sigma-models. The utility of
membrane sigma-models as a fundamental microscopic description of closed strings
in non-geometric flux backgrounds was originally suggested by~\cite{Halmagyi},
and further elucidated in~\cite{Mylonas:2012pg,Chatzistavrakidis:2015vka,Bessho:2015tkk,Chatzistavrakidis:2018ztm}.

The upshot is that the general expressions for the fluxes and Bianchi identities of a generic string compactification are in direct correspondence to the axioms of a Courant algebroid and to the generalized Wess-Zumino terms of Courant sigma-models. One may then wonder whether this ``triple point'' also exists for M-theory compactifications. The main purpose of this paper is to investigate this problem for the $SL(5)$ case of M-theory flux compactifications to seven dimensions. 

Our approach comprises two steps. In a first step, we consider the higher Courant bracket on the extended bundle $E_2=TM\oplus \mbox{\footnotesize$\bigwedge$}^2\,T^{\ast}M$ and identify its possible twists, which in the physical case where $\dim M=4$ turn out to be 
\be \label{eq:introfluxes}
G_{ijkl}\ ,\quad F_{ij}{}^{k} \ , \quad Q_{i}{}^{jkl} \qquad
\mbox{and} \qquad {\cal R}^{i,jklm}\ .
\ee 
 These are the higher counterparts of the more familiar set of NS--NS
 fluxes $H_{ijk}$, $F_{ij}{}^{k}$, $Q_{i}{}^{jk}$ and $R^{ijk}$
 encountered in string compactifications. The first two are associated
 to geometric compactifications of M-theory, while
 the last two are non-geometric fluxes~\cite{Hull:2006tp} sourced by exotic branes~\cite{Bakhmatov:2017les}. In particular, the last entry
 is the higher analogue of the locally non-geometric 3-vector $R$-flux, and presently it
 necessarily has the structure of a mixed-symmetry 5-vector of type
 $(1,4)$, in accord with \cite{Blair:2014zba}. Using the higher Courant bracket on $E_2$ we determine the general expressions for the fluxes \eqref{eq:introfluxes} and their corresponding Bianchi identities in terms of a vielbein, a 3-form and a 3-vector.

In a second step, we employ the construction of topological field
theories of AKSZ-type in four worldvolume dimensions (an open
threebrane), studied in \cite{Ikeda:2010vz,Ikeda:2012pv}. The
classical master equation for the corresponding threebrane sigma-model
yields a set of conditions, which are then used to define the
higher structure of a Lie algebroid up to homotopy \cite{Ikeda:2010vz}---see also
\cite{Gru,Gru2,Carow-Watamura:2016lob} for a somewhat more general construction. Such threebrane
sigma-models were already used in~\cite{Kokenyesi:2018ynq} to relate
AKSZ theories of M2-branes in topological M-theory to exceptional
generalized geometry and fluxes. Here we show that by
choosing the extended bundle $E_2$ in which the fields of the
sigma-model take values, accompanied by a projection to $SL(5)$
tensors, the local coordinate expressions for the axioms of a specific
Lie algebroid up to homotopy on $E_2$ reproduce the expressions for the
M-theory fluxes and their Bianchi identities for compactifications to
seven dimensions. Therefore we conclude that the same set of equations
underlies the $SL(5)$ M-theory fluxes and their Bianchi identities,
the gauge structure of a topological sigma-model in four worldvolume
dimensions, and the axioms of a specific Lie algebroid up to
homotopy. 

This paper is organized as follows. In Section~\ref{sec2} we briefly
recall the necessary background material on the higher Courant bracket
for extended bundles of the type $E_p=TM\oplus
\mbox{\footnotesize$\bigwedge$}^{p}\,T^{\ast}M$, and the construction
of a generalized metric for the physically relevant case $p=2$ which
accounts for $SL(5)$ exceptional generalized geometry. In
Section~\ref{sec3} we use the higher Courant bracket to determine the
general expressions for the fluxes and their Bianchi identities, and
explain how in the case $\dim M=4$ they indeed yield the correct
$SL(5)$ tensor structures. In Section~\ref{sec4}, topological
sigma-models on an open threebrane are discussed, together with their
relation to the structure of a Lie algebroid up to homotopy; we
further present some simple examples and their relation to M-theory
backgrounds. In Section~\ref{sec5} the relation between the general
M-theory fluxes and generalized Wess-Zumino terms of the topological
sigma-models is established. Section~\ref{sec6} contains a first step
towards the generalization of our results to the extended geometry of
exceptional field theory; here the dimensionality of the target space is extended from four to ten and the general expressions for the $SL(5)$ exceptional field theory fluxes up to the section condition are determined, though we have not yet found the appropriate extension of our threebrane sigma-model whose Wess-Zumino terms support the exceptional field theory fluxes. Finally, Section~\ref{sec7} contains our conclusions and an outlook toward some open problems related to our work.    

\section{Higher Courant Algebroids}
\label{sec2}

\subsection{Higher Courant Brackets}
\label{sec21}

We begin with a brief description of higher analogues of Courant
algebroids (see for example~\cite{Hagiwara,BZ,Zambon,Bi,Bouwknegt}), and
their uses in defining a generalized metric associated to exceptional
group structures on their underlying vector bundle \cite{HullEGG} which extends the corresponding construction in generalized complex geometry \cite{Hitchin,Gualtieri}. 

The starting point is a $d$-dimensional manifold $M$ and a vector bundle $E_p\to M$, which is the extension of its tangent bundle by $p$-forms,
\be  
E_p=TM\oplus \mbox{\footnotesize${\bigwedge}$}^p\,T^{\ast}M~,
\ee
where $p$ is a non-negative integer. For $p=1$ this is simply the vector bundle corresponding to a splitting of the exact sequence
\be
0\longrightarrow T^*M \xrightarrow{ \ \rho^\top \ } E_1 \xrightarrow{ \
	\rho \ } TM \longrightarrow 0 
\ee
which gives rise to an exact Courant algebroid, where $\rho^\top:T^*M\to E_1$ denotes the transpose of the anchor map $\rho$. 
For any $p\ge 1$, sections of the vector bundle $E_p$ correspond to a formal sum of vectors and $p$-forms, 
\be 
\Gamma(E_p) \ni A=X+\eta \qquad \mbox{with} \quad X\in \Gamma(TM) \ , \ \eta\in \Gamma(\mbox{\footnotesize$\bigwedge$}^p\,T^{\ast}M)~.
\ee 
The 
vector bundle $E_p$ can be endowed with a non-degenerate symmetric fiber pairing 
\be 
\langle\,\cdot\,,\,\cdot\,\rangle\,: E_p\times E_p\longrightarrow \mbox{\footnotesize$\bigwedge$}^{p-1}\,T^{\ast}M~,
\ee  
which is given in terms of a symmetrization of contractions between vectors and $p$-forms,
\be \label{higherbilinear}
\langle X+\eta, Y+\xi\rangle=\sfrac 12 \, (\iota_X\xi +\iota_Y\eta)~,
\ee 
resulting in a $(p-1)$-form.
In the special case $p=1$, this is a map to $C^{\infty}(M)$ and it
defines a metric with split signature $(d,d)$, which is invariant
under the continuous T-duality group $O(d,d)$. 

A binary operation on sections of $E_p$ can also be defined, either in terms of a higher Dorfman bracket, which we denote here as a circle product
\be \label{dorfman}
(X+\eta)\circ (Y+\xi)=[X,Y]+{\cal L}_{X}\xi-\iota_{Y} \dd\eta~,
\ee 
or in terms of its antisymmetrization, a higher Courant bracket 
\be \label{courant}
[X+\eta,Y+\xi]=[X,Y]+{\cal L}_{X}\xi-{\cal L}_Y\eta -\sfrac 12 \, \dd (\iota_X\xi-\iota_Y\eta)~.
\ee 
Here ${\cal L}_X$ denotes the Lie derivative along the vector field
$X$. For $p=1$ these are precisely the standard Dorfman and Courant
brackets, and they are formally given in terms of the same expressions for higher $p$. 

We further define a smooth bundle map $\rho: E_p\to TM$, which could be for instance the projection to the tangent bundle, such that the quadruple $(E_p,\langle\,\cdot\,,\,\cdot\,\rangle,[\,\cdot\,,\,\cdot\,],\rho)$ satisfies the following properties:
		\begin{itemize}
		\item \ Modified Jacobi identity:
		\be [[A,B],C]+\text{cyclic}(A,B,C)= \dd\,{\cal N}(A,B,C)~;\ee
		\item \ Homomorphism property:\ \be \rho [A,B]=[\rho(A),\rho(B)]~;\ee
		\item \ Modified Leibniz rule:
		\be [A,f\,B]=f\,[A,B]+\big(\rho(A)f\big)\,B- \dd f \w \langle A,B\rangle~;\ee
		\item \ Compatibility condition:  
		\be {\cal L}_{\rho(C)}\langle A,B\rangle = \langle [C,A]+\dd\langle C,A\rangle,B\rangle + \langle A,[C,B]+\dd\langle C,B\rangle\rangle ~,\ee
	\end{itemize} 
where the Nijenhuis operator is defined by
\be 
{\cal N}(A,B,C)=\sfrac 13\, \langle [A,B],C\rangle +\text{cyclic}(A,B,C)~,
\ee
for any 
$A,B,C\in\G(E_p)$ and $f\in C^{\infty}(M)$. The last
condition, expressing the compatibility between the pairing and the
bracket, is somewhat modified as compared to the $p=1$ case, in the sense that a Lie derivative appears on the left-hand side. Evidently, this condition reduces to the standard one for a Courant algebroid in the $p=1$ case. 

When the anchor map $\rho$ is the projection to $TM$, the higher
Courant bracket may be twisted by a closed $(p+2)$-form $H$ as 
\be 
[X+\eta,Y+\xi]_H=[X+\eta,Y+\xi]+\iota_X\iota_YH~.
\ee 
The twist yields an additional $p$-form term in the bracket. In string
theory, where the relevant structure is $p=1$, the twist is a closed
3-form; it is customarily identified with the NS--NS flux whose
de~Rham cohomology class is the \v{S}evera class which classifies the exact Courant algebroids over $M$. 
In that case, one can also consider more general twists, usually
denoted as $(H,f,Q,R)$ corresponding to a 3-form $H$, a vector-valued
2-form $f$, a bivector-valued 1-form $Q$ and a trivector $R$. The relevant bracket
is then the Courant-Roytenberg bracket and the underlying structure corresponds to a general Courant algebroid where the anchor is not simply the projection to the tangent bundle, or equivalently to a protobialgebroid \cite{dee3,K-S,Chatzistavrakidis:2015vka}. 
However, for $p>1$, the twisting is more restricted. For instance, the analogue of a twisted Poisson structure (which would be a twisted Nambu-Poisson structure) does not exist \cite{Bouwknegt}. 
Nevertheless, one can directly infer from the entries of the bracket what are the five possible additional types of twists that could in principle be considered apart from a $(p+2)$-form: a vector-valued 2-form, a $(p+1)$-vector-valued 1-form, a $p$-vector-valued $(p+1)$-form, a $2p$-vector-valued $p$-form, and a $(2p+1)$-vector. The precise structure of such twists will be explained in the ensuing sections.

For our purpose of investigating the flux content of M-theory for
four-dimensional compactification manifolds, giving rise to
dimensional reductions to seven dimensions, the relevant higher
structure is simply the one with $p=2$ which corresponds to extending
string charges for $p=1$ to M2-brane charges~\cite{HullEGG}. We shall
also find it necessary to formulate these higher Courant algebroids
more generally later on by replacing the tangent bundle $TM$ and the
cotangent bundle $T^*M$ by a Lie algebroid $E$ over $M$ and its dual $E^*$.
In order to investigate cases with $d>4$ the above setting is not
sufficient and additional ingredients, in the form of extended
bundles, are necessary to account for the charges of
higher-dimensional objects such as M5-branes and
MKK6-monopoles~\cite{HullEGG}. Therefore, we mostly focus on the special
case of $p=2$ and $d=4$, although some of our results will be valid
for any compactification dimension $d$ as we indicate.   

\subsection{$SL(5)$ Exceptional Generalized Geometry}

Let us recall that in the more familiar $p=1$ case, a generalized metric on $E_1$ that combines a 
Riemannian metric $g$ and a 2-form $B$ on $TM$ may be defined \cite{Gualtieri}. One way to do this is to start from the 
$O(d,d)$-structure that preserves the fiber pairing of the Courant
algebroid. The $O(d,d)$ T-duality transformations split into three types. The first corresponds to smooth bundle automorphisms 
$f:TM\to TM$ and their inverse transpose $f^{-\top}:T^{\ast}M\to
T^{\ast}M$, acting on a section of $E_1=TM\oplus T^*M$ as
$X+\eta\mapsto f(X)+f^{-\top}(\eta)$,
or in $GL(d)\subset O(d,d)$ matrix form
\be 
F\begin{pmatrix}
	X \\ \eta
\end{pmatrix}=\begin{pmatrix}
f & 0 \\ 0 & f^{-\top}
\end{pmatrix}\begin{pmatrix}
X \\ \eta
\end{pmatrix} \ .
\ee 
The second set of transformations are the $B$-transforms acting as $X+\eta \mapsto X+\eta+\iota_{X}B$, or
\be \label{eq:Btransform}
e^{B}\begin{pmatrix}
	X \\ \eta
\end{pmatrix}=\begin{pmatrix}
1 & 0 \\ B & 1
\end{pmatrix}\begin{pmatrix}
X \\ \eta
\end{pmatrix}~.
\ee 
The third type are the $\beta$-transforms which act via a bivector
$\beta$ on $T^*M$ as $X+\eta\mapsto X+\eta+\iota_{\eta}\beta$, or
\be 
e^{\b}\begin{pmatrix}
	X \\ \eta
\end{pmatrix}=\begin{pmatrix}
	1 & \b \\ 0 & 1
\end{pmatrix}\begin{pmatrix}
	X \\ \eta
\end{pmatrix}~.
\ee 
The geometric subgroup of
these $O(d,d)$-transformations preserving the Courant bracket is 
$GL(d)\ltimes \Omega^2_{\text{cl}}$, the semi-direct product of
diffeomorphisms of $M$ with 
closed 2-form transformations which act as bundle automorphisms
\eqref{eq:Btransform} preserving the pairing \eqref{higherbilinear}. 

Then an $O(d,d)$-covariant generalized metric ${\cal H}_1$ may be parametrized in terms of $g$ and $B$ as 
\be 
{\cal H}_1=\begin{pmatrix} g-B\,g^{-1}\,B & -B\,g^{-1} \\ 
	g^{-1}\,B & g^{-1}
\end{pmatrix}~,
\ee
which is a $B$-transform of the induced Riemannian metric $g\oplus g^{-1}$ on $TM\oplus T^*M$.
This is not the most general parametrization of ${\cal H}_1$, since
$O(d,d)$-transformations generate fractional linear transformations of
$g+B$, while field redefinitions allow for instance an expression of
the generalized metric in terms of a metric $\tilde g$ and a bivector
$\beta$ on $T^*M$~\cite{Andriot:2011uh}. The generalized metric for $p=1$ yields a reduction of the structure group $O(d,d)$ of $E_1$ to its maximal compact subgroup $O(d)\times O(d)$, and thus the moduli space of such reductions is the $d^2$-dimensional coset $O(d,d)/O(d)\times O(d)$, 

A similar prescription was followed in \cite{HullEGG} to define a
generalized metric for $p=2$ and $d=4$. In that case, the group of
transformations acting on $E_2=TM\oplus\mbox{\footnotesize$\bigwedge$}^2\,T^*M$ is the U-duality group $SL(5)$,
and the geometric subgroup preserving the higher Courant bracket is
the semi-direct product of diffeomorphisms of $M$ with closed 3-form
transformations $\Omega_{\rm cl}^3$. The action can be decomposed into
four types, in a similar way as before. The 15-dimensional $SL(4)$
subgroup acts separately on vectors, in the representation
$\mathbf{4}$, and on 2-forms, in the representation $\bf 6$, which combine into the antisymmetric representation ${\bf 10}$ of $SL(5)$. A 3-form $C \in \Gamma(\mbox{\footnotesize$\bigwedge$}^3\,T^{\ast}M)$ acts as $X+\eta\mapsto X+\eta+\iota_XC$ ($C$-transform) and a 3-vector ${\mit\Omega}\in \Gamma(\mbox{\footnotesize$\bigwedge$}^3\,TM)$ acts as $X+\eta\mapsto   X+\eta+\iota_{\eta}{\mit\Omega}$ (${\mit\Omega}$-transform). 
Both the 3-form and the 3-vector have four independent components in four dimensions. 
The matrix versions of these transformations are very similar to those
of the $p=1$ case above, thus we do not write them explicitly. Finally, a scaling transformation 
$X+\eta\mapsto \a^3\,X+\a^2\,\eta$ with $\a\in \R^{\times}$ guarantees
closure of the group action. 
As before, a generalized metric ${\cal H}_2$ can then be parametrized in terms of a metric $g$ and a 3-form $C$ on $TM$ in the form \cite{HullEGG}
 \be \label{exgm}
 {\cal H}_2=\begin{pmatrix} g+\sfrac 12\, C\,g^{-1}{\wedge} g^{-1}\,C &
   -\sfrac 12\, C\,g^{-1}{\wedge} g^{-1} \\[4pt]
 	-\sfrac 12\, g^{-1}{\wedge} g^{-1}\,C & \sfrac 12\, g^{-1}{\wedge}
        g^{-1}
 \end{pmatrix}~.
 \ee 
In the present case the structure group of $E_2$ is $SL(5)$, its
maximal compact subgroup is $SO(5)$, and the moduli space of
reductions by ${\cal H}_2$ is the corresponding 14-dimensional coset $SL(5)/SO(5)$.

\section{$SL(5)$ M-Theory Fluxes}
\label{sec3}

\subsection{Fluxes from the Higher Courant Bracket}
\label{sec31}

To determine the general form of the fluxes in M-theory for $7+4$ dimensions, where the four internal dimensions are compactified, we follow a strategy similar to the one suggested in \cite{Halmagyi} for non-geometric fluxes in generalized geometry. Recall that in generalized geometry the expressions for all the types of fluxes may be found upon acting with the twist operator
$e^{B}\,e^{\beta}$ (which is an element of $O(d,d)$) on the local
holonomic basis
spanned by $\partial_i=\frac\partial{\partial x^i}$ and $\dd x^i$ \cite{Blumenhagen:2012pc,Chatzistavrakidis:2013wra}. This gives
\begin{subequations} 
	\begin{align} \label{dxe1}
\partial_i \, &\xrightarrow{e^B\,e^{\beta}} \,
e_i:=\partial_i+B_{ij}\,\dd x^j~,\\[4pt]
\dd x^i \, &\xrightarrow{e^B\,e^{\beta}} \, e^i:=\dd x^i+\beta^{ij}\,\partial_j+\beta^{ij}\,B_{jk}\,\dd x^k=\dd x^i+\beta^{ij}\,e_j
~.\label{dxe2}
\end{align} 
\end{subequations}
Then, computing the \emph{untwisted} Courant brackets of the new basis, one obtains  
\bse \begin{align}
\label{courbasis1} [e_i,e_j]&=H_{ijk}\,e^k+F_{ij}{}^k\,e_k~,\\[4pt]
\label{courbasis2} {[}e_i,e^j]&=F_{ik}{}^j\,e^k+Q_i{}^{jk}\,e_k~,\\[4pt]
\label{courbasis3} {[}e^i,e^j]&=Q_{k}{}^{ij}\,e^k+R^{ijk}\,e_k~,
\end{align} \ese 
and the fluxes are identified with the generalized structure constants appearing on the right-hand
side. Their explicit expressions are
\bse \begin{align} \label{R}
H_{ijk}&=3\,\partial_{[i}B_{jk]}, \\[4pt]
F_{ij}{}^k&=\beta^{kl}\,H_{lij}~,\\[4pt]
Q_{k}{}^{ij}&=\partial_{k}\beta^{ij}
+\beta^{il}\,\beta^{jm}\,H_{lmk}~,\\[4pt]
R^{ijk}&=
3\,\beta^{[i\underline{l}}\,\partial_l\beta^{jk]} 
+\beta^{il}\,\beta^{jm}\,\beta^{kn}\,H_{lmn}~,
\end{align} \ese
where underlined indices do not participate in the antisymmetrization.
The corresponding Bianchi identities may be obtained by using the
Jacobi identity for this bracket \cite{Blumenhagen:2012pc}. 

For later reference, let us recall that in a general non-holonomic basis the above expressions take the form 
\bse \begin{align}
H_{abc}&=3\,\nabla_{[a}B_{bc]}, \label{flux3da}\\[4pt]
F_{ab}{}^c&=f_{ab}{}^{c}+\beta^{cd}\,H_{abd}~,\label{flux3db}\\[4pt]
Q_{a}{}^{bc}&=\partial_{a}\beta^{bc}+\beta^{bd}\, f_{ad}{}^c -\beta^{cd}\, f_{ad}{}^b
+\beta^{bd}\,\beta^{ce}\,H_{ade}~,\label{flux3dc}\\[4pt]
R^{abc}&=
3\,\beta^{[a\underline{d}}\,\nabla_d\beta^{bc]} 
+\beta^{ad}\,\beta^{be}\,\beta^{cf}\,H_{def}~, 
\label{flux3dd}\end{align}\ese 
where $\nabla_a$ is the covariant derivative with respect to a
vielbein $e_a=e_{a}{}^{i}\,\partial_i$, and its dual
$e^a=e^a{}_i\,\dd x^i$, acting as 
\be 
\nabla_{a}B_{bc}=\partial_{a}B_{bc}-{\mit\G}_{ab}{}^{d}\,B_{dc}-{\mit\Gamma}_{ac}{}^{d}\,B_{bd}~,
\ee
 and 
\be 
f_{ab}{}^c=2\,e^c{}_{j}\,e_{[a}{}^i\,\partial_ie_{b]}{}^j=:2\,{\mit\Gamma}_{[ab]}{}^{c}
\ee 
is the purely geometric torsion flux, appearing for example in string compactifications on twisted tori \cite{Hull:2005hk}.

Following the same prescription in the present case,{\footnote{Everything
    that follows holds for an arbitrary dimensionality $d$ of $M$, as indicated; however, the physically relevant case is only the one with $d=4$, specified in parentheses. For higher $d$, the physical case requires a different extension of the tangent bundle, as already mentioned.}} we consider a local basis $\{e_I\}, I=1,\dots,\frac{d\,(d+1)}{2}$ $(I=1,\dots,10)$ of sections of the vector bundle $E_2$, which may be split in a $d+\frac{d\,(d-1)}{2} $ ($4+6$) fashion as
\be 
\{e_I\}=\{e_i,e^{ij}\}~, \quad i,j=1,\dots,d \ (i,j=1,2,3,4)~,
\ee  
with $e^{ij}=-e^{ji}$.
Then we compute the untwisted higher Courant bracket 
\be 
[e_I,e_J]=T_{IJ}{}^K\,e_K\ ,
\ee 
and identify the corresponding M-theory fluxes with the local structure constants $T_{IJ}{}^{K}$. Explicitly, 
the brackets are generally given as 
\bse\begin{align} \label{agg1}
[e_i,e_j]&=G_{ijkl}\,e^{kl}+F_{ij}{}^m\,e_m~,\\[4pt] \label{agg2}
{[}e_i,e^{jk}]&=\widetilde F_{ilm}{}^{jk}\,e^{lm}+Q_i{}^{jkm}\,e_m~,\\[4pt] \label{agg3}
{[}e^{ij},e^{kl}]&=\widetilde Q_{mn}{}^{ij,kl}\,e^{mn}+R^{ij,kl,n}\,e_n ~.
\end{align}\ese  
Following the terminology of the string theory case, we refer to $G$,
$F$ and $\widetilde{F}$ as geometric fluxes and to
$Q$, $\widetilde Q$ and $R$ as non-geometric fluxes. Compared to the
string theory case there is a proliferation in that $F$ and $Q$ do not
repeat in two different brackets, but rather two apparently new fluxes
appear, $\widetilde F$ and $\widetilde Q$. We shall determine their
relation to the fluxes $F$ and $Q$ below, where we will see that they
are not independent. There is further a proliferation in the
index structure; for instance, the previously trivector flux $R$ now
becomes a (mixed-symmetry) tensor with five indices (a 5-vector,
but not fully antisymmetric---a completely antisymmetric tensor would anyway
vanish in four dimensions) and index structure $(2,2,1)$. Recalling
that previous studies of M-theory fluxes have revealed a
mixed-symmetry 5-vector of type $(1,4)$
\cite{Blair:2014zba,Bosque:2016fpi,Gunaydin:2016axc,Kupriyanov:2017oob,Lust:2017bgx,Lust:2017bwq},
it is natural to ask what is the relation between the two. We shall
return to this point after computing the explicit expressions for these fluxes. 

As in the string theory case, we begin with the coordinate basis of sections of $E_2=TM\oplus \mbox{\footnotesize$\bigwedge$}^2\,T^\ast M$, spanned by $\partial_i$ and $\sfrac 12\, \dd x^i\w\dd x^j$, and twist them by $SL(5)$ transformations corresponding to the action of a 3-form $C=\sfrac 16\, C_{ijk}\,\dd x^i\w\dd x^j\w\dd x^k$ and a 3-vector ${\mit\Omega}= \sfrac 16\, {\mit\Omega}^{ijk}\,\partial_i\w\partial_j\w\partial_k$ to get
\bse\begin{align} \label{xe1}
\partial_i \, &\xrightarrow{e^C\,e^{{\mit\Omega}}} \,
e_i:=\partial_i+\sfrac 12 \, C_{ijk}\,\dd x^j\w\dd x^k~,\\[4pt] \label{xe2}
\sfrac 12 \, \dd x^i\w\dd x^j \, &\xrightarrow{e^C\,e^{{\mit\Omega}}} \,
                                   e^{ij}:=\sfrac 12 \, \dd x^i\w\dd x^j +\sfrac 12\, {\mit\Omega}^{ijk}\, e_k
~.
\end{align}\ese 
We now compute the \emph{untwisted} higher Courant brackets of the new basis $\{e_i, e ^{ij}\}$  and comparing with \eqref{agg1}--\eqref{agg3} we formally obtain
\bse \begin{align}
\label{exflux1} G_{ijkl}&=4\,\partial_{[i}C_{jkl]}~,\\[4pt]
\label{exflux2} F_{ij}{}^m&=-\sfrac 12\,G_{ijkl}\,{\mit\Omega}^{klm}~,\\[4pt]
\label{exflux3} \widetilde F_{ilm}{}^{jk}&=\sfrac 12\,G_{ilmn}\,{\mit\Omega}^{njk}~,\\[4pt]
\label{exflux4} Q_i{}^{jkm}&=\sfrac 12\,\big(\partial_i{\mit\Omega}^{jkm}-\sfrac 12\,{\mit\Omega}^{jkn}\,G_{inps}\,{\mit\Omega}^{psm}\big)~,\\[4pt]
\label{exflux5} \widetilde Q_{mn}{}^{ij,kl}&=-\sfrac 14\,\big(\delta^l_{[m}\,\partial_{n]}{\mit\Omega}^{ijk}- \delta^k_{[m}\,\partial_{n]}{\mit\Omega}^{ijl}- \delta^j_{[m}\,\partial_{n]}{\mit\Omega}^{ikl}+ \delta^i_{[m}\, \partial_{n]}{\mit\Omega}^{jkl} \nonumber\\ &\hspace{2cm} +{\mit\Omega}^{ijp}\,G_{pp'nm}\,{\mit\Omega}^{p'kl}\big)~,\\[4pt]
\label{exflux6} R^{ij,kl,n}&=\sfrac12\,\hat\partial^{i[j}{\mit\Omega}^{kln]}-\sfrac12\,\hat\partial^{j[i}{\mit\Omega}^{kln]}-\sfrac12\,\hat\partial^{k[l}{\mit\Omega}^{ijn]}+\sfrac12\,\hat\partial^{l[k}{\mit\Omega}^{ijn]}\nonumber \\ &\hspace{2cm} -\sfrac 18\,{\mit\Omega}^{ijm}\,{\mit\Omega}^{klp}\,{\mit\Omega}^{rsn}\,G_{mprs}~,
\end{align}\ese
where we defined
$ \hat\partial^{ij}:={\mit\Omega}^{ijk}\,\partial_k$. These
expressions were derived without using the restriction $d=4$. One may directly observe that 
the fluxes $\widetilde F$ and $\widetilde Q$ are not independent from $F$ and $Q$, as there are the trace relations
\begin{align}
\label{ftildef} \widetilde{F}_{ijl}{}^{lk}&=F_{ij}{}^{k}~, \\[4pt]
\label{qtildeq} \widetilde{Q}_{im}{}^{jk,lm}&=\sfrac d4\,Q_i{}^{jkl}+\sfrac{d-4}{16}\,{\mit{\Omega}}^{jkp}\,G_{ipqr}\,{\mit{\Omega}}^{qrl}-\sfrac 14 \, \d_{i}^{[j}\,\partial_{n}{\mit{\Omega}^{k]ln}}+\sfrac 18\,\d_i^l\,\partial_{n}{\mit{\Omega}}^{jkn}~.
\end{align} 
For general $d$, the flux
$Q_i{}^{jkl}$ is not completely antisymmetric in its three vector indices.

Let us now specialize to the four-dimensional case. First, it is then inevitable that only the contracted parts of $\widetilde F$ and $\widetilde Q$ play a role, since they both have more than four indices in total. Specifically, the expression \eqref{qtildeq} for $\widetilde{Q}$ results in{\footnote{Note that the trace of the second term in $\widetilde{Q}$, namely ${\mit{\Omega}}^{jkp}\,G_{pqrs}\,{\mit{\Omega}}^{qrs}$, vanishes in four dimensions.}} 
\be
\widetilde{Q}_{im}{}^{jk,lm}=Q_i{}^{jkl}-\sfrac 12 \, \d_{i}^{[j}\,Q_{n}{}^{k]ln}+\sfrac 14\,\d_i^l\,Q_{n}{}^{jkn}~.
\ee 
Second, in four dimensions, the flux $Q_{i}{}^{jkl}$ is necessarily completely antisymmetric in its three vector indices, in agreement with expectations. Another important observation is that the last term in \eqref{exflux6} 
is identically zero in four dimensions for any antisymmetric 3-vector ${\mit\Omega}$.
Thus it is useful to define the $(1,4)$ mixed-symmetry combination 
\be 
{\cal R}^{i,jklm}=\sfrac12\,\hat{\partial}^{i[j}{\mit\Omega}^{klm]}~,
\ee 
which allows us to write the $R$-flux obtained by the higher Courant bracket in terms of the ${\cal R}$-flux as 
\be 
R^{ij,kl,n}={\cal R}^{i,jkln}-{\cal R}^{j,ikln}-{\cal R}^{k,lijn}+{\cal R}^{l,kijn}~.
\ee  
Thus unlike the string theory case where all indices of the trivector $R$-flux 
participate in the antisymmetrization, here the right-hand side contains terms with derivatives of the trivector ${\mit\Omega}$ where one index is outside the antisymmetrization. 
Studying all possibilities for index assignments, it turns out that in
all cases we necessarily end up with only one term{\footnote{The
    argument works as follows. Fix $i=i_0\in\{1,2,3,4\}$. If $i_0=j$ then obviously the $R$-flux vanishes; if $i_0=k$, then $R^{ij,kl,n}=\,\hat{\partial}^{i_0[j}{\mit\Omega}^{i_0ln]}$, and similarly if $i_0=l$; if $i_0=n$ then $R^{ij,kl,n}=\sfrac 12\,\hat{\partial}^{i_0[j}{\mit\Omega}^{kli_0]}$. If $i_0\ne j,k,l,n$, then fix $j=j_0$ and repeat the argument.}} and the $R$-flux is a $(1,4)$ mixed-symmetry tensor, in agreement with expectations. 
 We discuss this point further in the extended case of Section~\ref{sec6}, where we shall compare with previous results in the literature.
  For the time being, we can already conclude that by studying the higher Courant bracket, the types of geometric and non-geometric fluxes which arise in the four-dimensional case are 
 \be 
 G_{ijkl}~, \quad F_{ij}{}^k~, \quad Q_{i}{}^{jkl} \qquad \mbox{and} \qquad  {\cal R}^{i,jklm}~.
 \ee 
Finally, the expressions above are written in the holonomic frame. The
corresponding expressions in a non-holonomic frame may be obtained
similarly, by considering a vielbein $e_a$ and
$e^{ab}=\frac12\,e^a\wedge e^b$. We present them only for the relevant fluxes in the four-dimensional case, which are 
\bse \begin{align} 
G_{abcd}&=4\,\nabla_{[a}C_{bcd]}~,
\label{flux4da}\\[4pt]
F_{ab}{}^{c}&=f_{ab}{}^{c}-\sfrac 12\, G_{abde}\,{\mit\Omega}^{dec}~,
\label{flux4db}\\[4pt]
Q_{a}{}^{bcd}&=\sfrac 12\, \big(\partial_{a}{\mit\Omega}^{bcd}+3\,{\mit\Omega}^{e[bc}\,f_{ae}{}^{d]}
-\sfrac 12\, {\mit\Omega}^{def}\,\d_{a}^{[b}\,f_{ef}{}^{c]}-\sfrac 12\, {\mit\Omega}^{e[bc}\,G_{aefg}\,{\mit\Omega}^{d]fg}\big)~,
\label{flux4dc}\\[4pt]
R^{ab,cd,e}&=\sfrac12\,\widehat\nabla{}^{a[b}{\mit\Omega}^{cde]}-\sfrac12\,\widehat\nabla{}^{b[a}{\mit\Omega}^{cde]}-\sfrac12\,\widehat\nabla{}^{c[d}{\mit\Omega}^{abe]}+\sfrac12\,\widehat\nabla{}^{d[c}{\mit\Omega}^{abe]}~,
\label{flux4dd}\end{align}\ese 
where $\widehat\nabla{}^{ab}={\mit\Omega}^{abc}\,\nabla_c$, and we used the definitions and facts explained before. Once more, one should keep in mind that the $R$-flux effectively contains only one term.\footnote{Upon dimensional reduction over the M-theory circle, the four-dimensional fluxes \eqref{flux4da}--\eqref{flux4dd} reduce to the NS--NS fluxes \eqref{flux3da}--\eqref{flux3dd} in three dimensions. In higher dimensions this is not generally true, as the $G$-flux can reduce to either the NS--NS $H$-flux or the 4-form RR flux depending on whether or not the 4-form $G$ has a leg along the M-theory circle.}

\subsection{Bianchi Identities}
\label{sec32} 

Bianchi identities for the above fluxes can be obtained upon calculating the Jacobiator for the higher Courant bracket using the basic property
\be\label{aaa}
[[A,B],C]+{\rm cyclic}(A,B,C)=\sfrac 13\,{\dd}\big(\langle[A,B],C\rangle+{\rm cyclic}(A,B,C)\big)~.\ee
To calculate the Jacobiator it is useful to write down the pairing $\langle A,B\rangle$ for the basis elements \eqref{xe1} and \eqref{xe2}, which reads as
\bse\begin{align}
\langle e_i,e_j\rangle&=0~,\\[4pt]
\label{bilinear2}\langle e_i,e^{jm}\rangle&=\sfrac 12\,\delta^{[j}_i\,\dd x^{m]}~,\\[4pt]
\langle e^{ij},e^{mn}\rangle &=\sfrac 12\,{\mit\Omega}^{[ijm}\,\delta^{n]}_l\,\dd x^{l}~.
\end{align}\ese
Using these expressions, the modified Jacobi identity containing only the basis elements $e_i$,
\be
[[e_i,e_j],e_m]+{\rm cyclic}(i,j,m)-\sfrac 13\,{\dd}\big(\langle[e_i,e_j],e_m\rangle+{\rm cyclic}(i,j,m)\big)=0~,\ee
gives directly the Bianchi identities
\bse
\begin{align}
\label{exbi1} \partial_{[m}G_{ijkl]}&=-\sfrac 35\, G_{np[ij}\,\widetilde F_{m]kl}{}^{np}-\sfrac 35\, F_{[ij}{}^n\,G_{m]nkl}~,\\[4pt]
\label{exbi2} \partial_{[m}F_{ij]}{}^l-\sfrac 1{3}\,\hat\partial^{lk}G_{ijmk}&=-G_{nk[ij}\,Q_{m]}{}^{nkl}-F_{[ij}{}^k\,F_{m]k}{}^{l}~.
\end{align}\ese
As in Section~\ref{sec31}, these formulas hold in any dimension
$d$. However, for the physically interesting case $d=4$, the first identity becomes algebraic, since the left-hand side is identically zero (five antisymmetrized indices in four dimensions), and taking into account the trace relation \eqref{ftildef} between $\widetilde F$ and $F$, the result is 
\be \label{eq:GF0}
G_{n[lij}\,F_{mk]}{}^{n}=0~.
\ee 
The rest of the Jacobi identities, involving different combinations of
the basis elements $e_i$ and $e^{ij}$, give six additional Bianchi
identities for all fluxes which read as
\bse \begin{align}
\label{exbi3}&3\,\partial_{[i}\widetilde F_{jp]r}{}^{mn}
-\delta^{[n}_{[r}\,\partial_{p]} F_{ij}{}^{m]}+\sfrac 12\,\hat\partial^{mn}G_{ijpr}+{\mit\Omega}^{ks[m}\,\delta^{n]}_{[p}\,\partial_{r]}G_{ijks}\nn\\
&\hspace{1cm} = G_{ijkl}\,\widetilde Q_{pr}{}^{kl,mn}+ F_{ij}{}^k\,\widetilde F_{kpr}{}^{mn}+2\,\widetilde F_{kl[i}{}^{mn}\,\widetilde F_{j]pr}{}^{kl}+2\,Q_{[i}{}^{mnk}\,G_{j]kpr}~,\\[4pt]
&2\,\partial_{[i}Q_{j]}{}^{mnp}+\sfrac 12\,\hat\partial^{mn}F_{ij}{}^p-\sfrac 12\,\hat\partial^{p[n} F_{ij}{}^{m]}-\sfrac 12\,\hat\partial^{lp}\widetilde F_{ijl}{}^{mn}+\sfrac 12\,{\mit\Omega}^{kl[m}\,\hat\partial^{n]p}G_{ijkl}\nn\\
&\hspace{1cm} = G_{ijkl}\,R^{kl,mn,p}+3\, F_{[ij}{}^k\,Q_{k]}{}^{mnp}+2\,\widetilde F_{kl[i}{}^{mn}\,Q_{j]}{}^{klp}~,\\[4pt]
&3\,\partial_{[i}\widetilde Q_{st]}{}^{jk,mn}-\hat\partial^{jk}\widetilde F_{ist}{}^{mn}-(\{jk\}\to\{mn\})\nn\\
&\hspace{1cm}=2\,\widetilde F_{ipr}{}^{jk}\,\widetilde
  Q_{st}{}^{pr,mn}-\widetilde F_{ist}{}^{pr}\,\widetilde
  Q_{pr}{}^{jk,mn}+2\,Q_i{}^{jkp}\,\widetilde F_{pst}{}^{mn}+
  R^{jk,mn,p}\,G_{pist} \nn\\
&\hspace{1cm}\quad-(\{jk\}\to\{mn\})~,\\[4pt]
&\partial_iR^{jk,mn,s}-\hat\partial^{jk}Q_{i}{}^{mns}-\hat \partial^{ps}\widetilde Q_{ip}{}^{jk,mn}
-(\{jk\}\to\{mn\})\nn\\
&\hspace{1cm}=2\,\widetilde F_{ipr}{}^{jk}\,R^{pr,mn,s}+
  F_{pi}{}^{s}\,R^{jk,mn,p}+2\,Q_i{}^{jkp}\,Q_{p}{}^{mns}- \widetilde
  Q_{pr}{}^{jk,mn}\,Q_{i}{}^{prs}\nn\\
&\hspace{1cm}\quad-(\{jk\}\to\{mn\})~,\\[4pt]
&\partial_{[t}R^{ij,kl,[m}\,\delta^{n]}_{s]}+\sfrac 3{4}\,\hat\partial^{mn}\widetilde Q_{st}{}^{ij,kl}+{\mit\Omega}^{[prm}\,\delta^{n]}_{[s}\,\partial_{t]}\widetilde Q_{pr}{}^{ij,kl}+{\rm cyclic} (ij,kl,mn) \nn\\
&\hspace{1cm}=R^{ij,kl,p}\,\widetilde F_{pst}{}^{mn}+\widetilde
  Q_{pr}{}^{ij,kl}\,\widetilde Q_{st}{}^{pr,mn}+\widetilde
  Q_{pr}{}^{ij,kl}\,\delta_{[s}^{[n}\,\big(Q_{t]}{}^{prm]}+\sfrac
  14\,{\mit\Omega}^{pr\underline{a}}\,G_{t]abc}\,{\mit\Omega}^{m]bc}\big) \nn
       \\
&\hspace{1cm}\quad+{\rm cyclic} (ij,kl,mn) ~,\\[4pt]
\label{exbi8}&\hat\partial^{mn}R^{ij,kl,q}+\sfrac 23\,\hat\partial^{pq}R^{ij,kl,[m}\,\delta^{n]}_{p}+\sfrac 23\,{\mit\Omega}^{[prm}\,\hat\partial^{n]q}\widetilde Q_{pr}{}^{ij,kl}+{\rm cyclic} (ij,kl,mn) \nn\\
&\hspace{1cm}=2\,R^{ij,kl,p}\,Q_{p}{}^{mnq}+2\,\widetilde Q_{pr}{}^{ij,kl}\,R^{pr,mn,q}+\sfrac 23\,\widetilde Q_{{ pr}}{}^{ij,kl}\,{\mit\Omega}^{dq[n}\,\big(Q_d{}^{prm]}+\sfrac 14\, {\mit\Omega}^{pr\underline{a}}\,G_{dabc}\,{\mit\Omega}^{m]bc}\big)\nn\\
&\hspace{1cm}\quad+{\rm cyclic} (ij,kl,mn) ~.
\end{align}\ese
Again these expressions are given in the holonomic frame. The Bianchi identities in a non-holonomic frame may be found in the same way.

\section{Threebrane Sigma-Models and Homotopy Algebroids}
\label{sec4}

\subsection{Overview}
\label{sec41}

Let us begin with an explanation of which precise question we aim at answering in the following. 
To this end let us go back to the string theory case, and the expressions for fluxes and Bianchi identities there. As already mentioned, they may be determined with the same approach as the one we used in Section~\ref{sec3} for M-theory. The essential point we would like to recall is as follows. First we note that the fluxes and Bianchi identities in the holonomic frame for generalized geometry may be written in the compact form
\begin{align}
\label{ca2} \rho^i{}_{I}\,\partial_i\rho^j{}_J-\rho^i{}_J\,\partial_i\rho^j{}_I-\eta^{KL}\,\rho^j{}_K\,T_{LIJ}&=0~,\\[4pt]
\label{ca3} 4\,\rho^i{}_{[L}\,\partial_iT_{IJK]}+3\,\eta^{MN}\,
T_{M[IJ}\,T_{KL]N}&=0~,
\end{align}
where indices $i,j,\dots$ run over $1,\dots,d$ while $I,J,\dots$ run
through $1,\dots,2d$. Here $\rho^i{}_I=(\d^i{}_j,\beta^{ij})$,
$\eta_{IJ}$ is the $O(d,d)$-invariant metric, and the 3-form $T_{IJK}$
corresponds to the four fluxes $H_{ijk}$, $F_{ij}{}^{k}$, $Q_i{}^{jk}$
and $R^{ijk}$ depending on the index position.
We supplement these equations with a third one, 
\be \label{ca1}
\eta^{IJ}\,\rho^i{}_{I}\,\rho^j{}_J=0~,
\ee 
which is identically satisfied as long as $\beta^{(ij)}=0$, namely
$\beta^{ij}$ are components of an antisymmetric 2-vector (not
necessarily a Poisson bivector). These three equations may be interpreted in three different but related ways:
\begin{itemize}
	\item As the fluxes and Bianchi identities in the NS--NS
          sector of general string theory compactifications.
	\item As the local form of the axioms of a Courant algebroid on $E_1$ \cite{Ikeda:2012pv}.
	\item As the conditions arising from the classical master equation in the BV--BRST quantization of generalized Wess-Zumino terms, or equivalently as the conditions for gauge invariance and on-shell closure of gauge transformations for the Courant sigma-model \cite{Ikeda:2002wh}.
\end{itemize}
Read differently, the above statements say the following: given a
Courant algebroid, one can uniquely write down a membrane sigma-model
which gives the BV--BRST action for Wess-Zumino terms which appear as
fluxes in string theory compactifications. This statement is based on
\cite{Ikeda:2002wh,Park:2000au,Hofman:2002jz,dee1,dee2}. In
particular, in~\cite{dee2} the precise relation with topological
sigma-models of AKSZ-type \cite{Alexandrov:1995kv} is demonstrated. 

Moving one worldvolume dimension higher, from the closed string to the closed M2-brane, we have already found the expressions for the fluxes and their Bianchi identities in Section~\ref{sec3}. This allows us to pose the following question: given a higher analogue of an exact Courant algebroid, can one write down uniquely a threebrane sigma-model which gives the BV--BRST action for generalized Wess-Zumino terms that appear as fluxes in M-theory  compactifications? 

In other words, we would like to have a set of expressions much like
\eqref{ca2}--\eqref{ca1}, which can be interpreted again in three
different ways: as M-theory fluxes and Bianchi identities, as the
local form of the properties of the higher Courant bracket, and as
conditions that guarantee the classical master equation for some
topological threebrane sigma-model (or, equivalently, its gauge
invariance and on-shell closure of gauge transformations). We address
this question in Section~\ref{sec5} below. In the present section we first discuss the construction of threebrane sigma-models as the higher analogue of Courant sigma-models. 

\subsection{Topological Threebrane Sigma-Models}
\label{sec42}

The threebrane sigma-model that corresponds to the topological AKSZ
theory for open threebranes, or in other words the BV--BRST action for a
4-form Wess-Zumino term, was contructed explicitly in
\cite{Ikeda:2010vz}. We shall not review the full construction
here,{\footnote{Along the usual lines of the BV--BRST formalism this
    consists in considering the BRST symmetry by replacing gauge
    parameters by ghosts, higher gauge parameters by
    ghosts-for-ghosts, and introducing the corresponding antifields
    required of the BV formalism. In the full space of fields, ghosts
    and antifields, an antibracket is defined and a BV--BRST action in
    constructed in terms of superfields on a graded manifold, which is
    required to satisfy the classical master equation imposing
    BRST-invariance.}} but instead we consider just the zero-ghost topological action for the AKSZ topological sigma-model in four dimensions, which will be enough to make our main point. Its general form reads as
\begin{align}
S[X,\a,A,F]&=\int_{\S_4}\, \big(F_i\w\dd X^i-\a_I\w\dd A^I+\rho^i{}_{
             I}(X)\,F_i\w A^I\,+\sfrac 12\, S^{IJ}(X)\,\a_I\w \a_J \label{qp3}\\
&\hspace{1cm}\qquad+ \,\sfrac 12\, T^{I}{}_{JK}(X)\,\a_I\w A^J\w A^K
+\sfrac 1{4!}\,G_{IJKL}(X)\,A^I\w A^J\w A^K\w A^L\big)~. \nn
\end{align}
Let us spell out the ingredients in this sigma-model action. We have a theory of maps from a threebrane worldvolume $\S_4$ to a $d$-dimensional target space $M$ (we shall mostly consider $d=4$ later, but the discussion here holds for any $d$), 
\bea
X=(X^i):\S_4\longrightarrow M \qquad \mbox{with} \quad i=1,\dots , d~,
\eea
and $F\in {\Omega}^{3}(\S_4,X^{\ast}T^{\ast}M)$ is an auxiliary worldvolume 
3-form taking values in the pullback of the cotangent bundle of $M$ by the map $X$.  In addition, there is a worldvolume 1-form $A\in{\Omega}^1(\S_4,X^{\ast}E)$ and a worldvolume 2-form 
$\a\in{\Omega}^2(\S_4,X^{\ast}E^{\ast})$. They take values in the
pullback of a vector bundle $E\to M$ and its dual $E^*\to M$ respectively. For example this could be the tangent bundle, but not necessarily so (see below). Here $I$ is a bundle index when a basis of local sections $\{e_I\}$ of $E$ is 
chosen, with corresponding dual basis $\{e^I\}$ for $E^*$, while $\rho^i{}_I$ are the components of an anchor map
$\rho: E \to TM$, and $S^{IJ}$ is symmetric in its two bundle indices,
which defines a symmetric bilinear pairing on sections of $E^*$
(possibly degenerate). Finally, $T^{I}{}_{JK}$ are structure constants of an antisymmetric bracket on sections of $E$ and $G_{IJKL}$ is a generalized 4-form on $E$.  
The quantities $S$, $T$ and $G$ are all functions of $X(\s)$, where
$\s^{\a}$  are the local coordinates of the threebrane worldvolume, and they will be understood as generalized Wess-Zumino couplings in the present setting. 

There is a hierarchy of structures and fields exhibited in the
following table:
\small
\begin{center}\boxed{
		\begin{tabular}{cccccc}
			\underline{$\dim\S$} & \underline{AKSZ $\sigma$-model} & \underline{0-forms} & \underline{1-forms} & \underline{2-forms} & \underline{3-forms}
			\\[4pt]
			2 & Poisson & $X^i$ & $F_i\in\G(X^{\ast}T^{\ast}M)$ & --- & --- 
			\\[4pt]
			3 & Courant & $X^i$ & $A^I\in \G(X^{\ast}E)$ & $F_i\in\G(X^{\ast}T^{\ast}M)$ & --- 
			\\[4pt]
			4 & Threebrane & $X^i$ & $A^I \in\G(X^{\ast}E)$ & $\a_I\in \G(X^{\ast}E^{\ast})$ & $F_i\in \G(X^{\ast}T^{\ast}M)$
	\end{tabular}}
\end{center}
\normalsize
Specifically, the AKSZ theory in two dimensions corresponds to the
BV--BRST action for the Poisson sigma-model (the first order
formulation of the topological bosonic string $B$-field amplitude), in three dimensions to
the Courant sigma-model and in four dimensions to the threebrane sigma-model that
we discuss
here. This is of course part of a semi-infinite staircase of (higher)
geometric structures and topological sigma-models. More details may be found for example in the review \cite{Ikeda:2012pv}.

The action \eqref{qp3} comes with a host of conditions stemming from the classical master equation---or, equivalently, from gauge invariance. 
Recall that in the two-dimensional case these conditions are
equivalent to the vanishing of the Schouten-Nijenhuis bracket for a bivector
field, which is the condition for a Poisson structure on $M$ or the Lie algebroid axioms for
the cotangent bundle $T^*M$ equipped with the Koszul-Schouten bracket, while in the three-dimensional case they are equivalent to the axioms of a Courant algebroid.
In the four-dimensional case the conditions 
found in \cite{Ikeda:2010vz,Gru} define a higher algebroid structure,
called a Lie algebroid up to homotopy in \cite{Ikeda:2010vz} or more generally an
$H$-twisted Lie algebroid in \cite{Gru} (see also Refs. \cite{Gru2,Carow-Watamura:2016lob}). Instead of providing the geometric axioms defining such a structure, we find it more 
illuminating at this stage to reside on the equivalent local coordinate conditions imposed by the classical master 
equation; the two approaches anyway reflect the same structure. 
The action \eqref{qp3} is invariant under the gauge transformations parametrized by scalar, 1-form and 2-form gauge parameters $\epsilon$, $\zeta$ and $t$:
\bse\begin{align}
\label{gt1} \delta X^i&=-\rho^i{}_I\, \epsilon^I~,\\[4pt]
\label{gt2} \delta A^I&=\dd\epsilon^I+S^{IJ}\,\zeta_J-T^I{}_{JK}\,A^J\,\epsilon^K~,\\[4pt]
\label{gt3} \delta \alpha_I&=\dd \zeta_I+\rho^{i}{}_{I}\,t_i+T^J{}_{IK}\,\zeta_J\w A^K+T^J{}_{IK}\,\alpha_J\,\epsilon^K+\sfrac 12\, G_{IJKL}\,\epsilon^J\,A^K\w A^L~,\\[4pt]
\delta F_i&=-\dd t_i+\partial_i\rho^j{}_I\,\big(\epsilon^I\,F_j+t_{j}\w A^{I}\big)-\partial_iT^J{}_{LI}\,\epsilon^I\,\alpha_J\w A^L \nn\\
\label{gt4} &\quad\,-\sfrac 16\, \partial_i G_{IJKL}\,\epsilon^I\,A^J\w A^K\w A^L +\sfrac 12\,\partial_iT^I{}_{JK}\,\zeta_I\w A^J\w A^K+\partial_i S^{IJ}\,\zeta_I\w\alpha_J~,
\end{align}\ese
provided the following conditions are met:
\bse\begin{align} 
\label{c1} \rho^i{}_I\,S^{IJ}=0~,\\[4pt]
\label{c2} \rho^{i}{}_I\,\partial_iS^{JK}+S^{LJ\,}T^K{}_{ IL}+S^{LK}\,T^J_{IL}=0~,\\[4pt]
\label{c3} \rho^i{}_I\,\partial_i\rho^j{}_J-\rho^i{}_J\,\partial_i\rho^j{}_I-\rho^j{}_K\,T^K{}_{ IJ}=0~,\\[4pt]
\label{c4} 3\,\rho^i{}_{[I}\,\partial_iT^J{}_{ KL]}+S^{JM}\,G_{KLIM}-3\,T^J{}_{ M[K}\,T^M{}_{ LI]}=0~,\\[4pt]
\label{c5} \rho^i{}_{[I}\,\partial_iG_{JKLM]}+T^N{}_{ [IJ}\,G_{KLM]N}=0~.
\end{align}\ese
Therefore, given a set of structure functions that solve these conditions, the
corresponding topological threebrane sigma-model 
is uniquely determined. One can then reconstruct the anchor, 
bracket and 4-form on the vector bundle $E$, as well as the pairing on
its dual $E^*$, as derived brackets \cite{Ikeda:2010vz}.

These relations define a homotopy deformation of a Lie algebroid on
$E$: Setting $S=G=0$ reduces the conditions \eqref{c1}--\eqref{c5} to
the usual axioms of a Lie algebroid, with the remaining non-trivial
identities \eqref{c3} and \eqref{c4} corresponding to the homomorphism
property, Leibniz rule and Jacobi identity for the anchor map and
bracket on $E$. Generally, the bracket $[\,\cdot\,,\,\cdot\,]$ on $E$
extends to give a higher analog of the (twisted) Courant bracket on
sections of $E\oplus\mbox{\footnotesize$\bigwedge$}^2E^*$, which can
again be computed as a derived bracket and for $S=0$ is given by~\cite{Ikeda:2010vz}
\be\label{eq:derivedbracket}
[s_1+\gamma_1,s_2+\gamma_2]_G = [s_1,s_2] +{\cal L}_{s_1}\gamma_2-{\cal L}_{s_2}\gamma_1-\sfrac12\,\dd_E(\iota_{s_1}\gamma_2-\iota_{s_2}\gamma_1) + \iota_{s_1}\iota_{s_2}G \ ,
\ee
where $s_1,s_2\in\Gamma(E)$ and
$\gamma_1,\gamma_2\in\Gamma(\mbox{\footnotesize$\bigwedge$}^2E^*)$. Here ${\cal
  L}_{s_i}=\dd_E\,\iota_{s_i}+\iota_{s_i}\,\dd_E$ and $\dd_E:\Gamma(\mbox{\footnotesize$\bigwedge$}^pE^*)\to\Gamma(\mbox{\footnotesize$\bigwedge$}^{p+1}E^*)$
is the usual Lie algebroid differential defined by
\be
(\dd_E\,\omega)_{I_0I_1\cdots
  I_p}=\rho^i{}_{[I_0}\,\partial_i\,\omega_{I_1\cdots
  I_p]}+T^J{}_{[I_0I_1}\,\omega_{I_2\cdots I_{p-1}]J}
\ee
for
$\omega=\frac1{p!}\, \omega_{I_1\cdots I_p}\,e^{I_1}\wedge\cdots \wedge
e^{I_p}$. Then \eqref{c4} (with $S=0$) is the nilpotency condition
$\dd_E^2=0$, while \eqref{c5} is just the closure condition
$\dd_E G=0$ which states that the twisting 4-form $G$ represents a
class in the degree~4 Lie algebroid cohomology of $M$; this classifies
the Lie algebroids up to homotopy over $M$ with $S=0$~\cite{Ikeda:2010vz}.

The relation of the threebrane sigma-model to the higher Courant bracket may also be established as follows. First recall that the generalized Wess-Zumino term for Courant sigma-models is 
\be \label{wztermCA}
\int_{\S_3}\,\sfrac 1{3!}\,X^{\ast}(T_{IJK})\,A^{I}\w A^{J}\w A^{K}=\int_{\S_3}\,\sfrac 1{3}\,X^{\ast}\big(\langle e_{I},[e_J,e_K]\rangle\big)\,A^{I}\w A^{J}\w A^{K}~,
\ee  
where the components $T_{IJK}$ are directly related to the twisted Courant-Roytenberg bracket in a local basis $\{e_I\}$ of $E_1=TM\oplus T^{\ast}M$ as 
$ 
T_{IJK}=2\,\langle e_{I},[e_J,e_K]\rangle$~\cite{dee2}.{\footnote{We indicated explicitly the pullback by $X$ here, in order to avoid confusion, in contrast to the customary short-hand notation $\sfrac 13\, \langle A,[A,A]\rangle$ for the integrand in \eqref{wztermCA} used e.g. in~\cite{Chatzistavrakidis:2018ztm}; the latter notation should be treated with caution, since the fields $A$ live in the pullback bundle $X^{\ast}E_1$, which unlike $E_1$ itself is not endowed naturally with a Courant algebroid structure. Thus the notation $[A,A]$ can be misleading, since no such bracket is defined on $X^{\ast}E_1$, and we refrain from using it here.}} In a similar fashion, the last two terms in the action \eqref{qp3} may be written as 
\bse 
\begin{align}
\sfrac 12 \,T^{I}{}_{JK}(X)\,\a_{I}\w A^{J}\w A^{K}& =X^{\ast}\big(\langle e^{I},[e_{J},e_{K}]\rangle\big)\,\a_{I}\w A^{J}\w A^{K}~,
\\[4pt]
\label{GWZ} \sfrac 14 \,G_{IJKL}(X)\, A^{I}\w A^{J}\w A^{K}\w A^{L}&=X^{\ast}\big(\langle e_{I},\langle e_J,[e_K,e_L]\rangle\rangle\big)\, A^{I}\w A^{J}\w A^{K}\w A^{L}~,
\end{align}
\ese 
where as before $\{e_I\}$ and $\{e^{I}\}$ are local bases of sections
of $E$ and $E^{\ast}$ respectively, the bracket is the higher Courant
bracket on $E\oplus\mbox{\footnotesize$\bigwedge$}^2E^*$ and the bilinear form
corresponds to symmetric contraction as in
\eqref{higherbilinear}:{\footnote{In \eqref{GWZ} the two bilinear
    forms are in principle different: the first (``inner'') bilinear
    form is defined on $E\oplus \mbox{\scriptsize$\bigwedge$}^2 E^*$, while the
    second (``outer'') bilinear form is the canonical pairing between the vector bundle $E$ and its dual $E^*$. Since in both cases the corresponding contractions are understood from the context, we refrain from establishing a separate notation for these two operations.}}
\be
\langle s_1+\gamma_1, s_2+\gamma_2\rangle=\sfrac 12\, (\iota_{s_1}\gamma_2 +\iota_{s_2}\gamma_1) \ .
\ee
Thus the higher geometric operations introduced above and in Section~\ref{sec21} directly dictate the generalized Wess-Zumino terms in the threebrane sigma-model. This will be further exemplified in a number of examples below. 

Therefore, the question we posed in Section \ref{sec41} may be rephrased as follows: What is the relation of the conditions \eqref{c1}--\eqref{c5} to the fluxes and Bianchi identities that we found in Section \ref{sec3}? Before we delve into the answer, we first discuss some (known and new) characteristic examples for this structure. 

\subsection{Examples}
\label{sec43}

\paragraph{Homotopy tangent algebroids.}

The simplest possibility is to choose $E=TM$ with the usual Lie
bracket of vector fields. Then the bundle index $I$ is identified with the
coordinate index $i$, in a local basis $\{e_i\}$ of the tangent
bundle. The worldvolume 1-form $A^i=A^{i}_{\a}\,\dd\s^{\a}$ is valued
in the pullback of the tangent bundle $X^{\ast}TM$ over $\S_4$ and the
worldvolume 2-form $\a_i=\sfrac 12\, \a_{i \, \a\b}\,\dd\s^{\a}\w\dd\s^{\b}$ is valued in the pullback of the cotangent bundle $X^{\ast}T^{\ast}M$.  

We begin with an analysis of the conditions \eqref{c1}--\eqref{c5} that define a Lie algebroid up to homotopy. The condition \eqref{c1} reads as 
\be 
\rho^i{}_k\,S^{kj}=0~. 
\ee 
If the pairing $S^{ij}$ is non-degenerate, this implies $\rho^i{}_j=0$. This is a legitimate option, especially in the case that the base manifold $M$ is a point and the algebroid structure is reduced to an algebra. Such cases were examined in \cite{Ikeda:2010vz}. For our purposes, it is more interesting to consider instead the case that the anchor is non-degenerate, in which case one concludes that 
\be 
S^{ij}=0~.
\ee  
Then \eqref{c1} and also \eqref{c2} are satisfied
automatically. Since $\rho$ is non-degenerate, with inverse
$\rho_i{}^j$, the condition \eqref{c3} requires that 
\be \label{geometricflux}
T^{i}{}_{jk}=2\,\rho_l{}^{i}\,\rho^{m}{}_{[j}\, \partial_{\underline{m}}\,\rho^{l}{}_{k]}~.
\ee  
Finally, the relations \eqref{c4} and \eqref{c5} resemble Bianchi identities and they explicitly read as 
\bse
\begin{align}
\rho^{l}{}_{[i}\,\partial_{\underline{l}}T^{j}{}_{mn]}-T^{j}{}_{l[m}\,T^{l}{}_{ni]}&=0~,\\[4pt]
\rho^{n}{}_{[i}\,\partial_{\underline{n}}G_{jklm]}+T^{n}{}_{[ij}\,G_{klm]n}&=0~.
\end{align} \ese
We distinguish two particularly interesting cases below.

\paragraph{$\boldsymbol G$-flux.}

Choose the anchor to be the projection to the tangent bundle, namely
$\rho=\text{id}$, or more explicitly
$\rho^{i}{}_{j}=\d^{i}{}_{j}$. Then \eqref{c3} implies immediately
that $T^{i}{}_{jk}=0$, which automatically satisfies
\eqref{c4}; in this case the derived
bracket \eqref{eq:derivedbracket} reproduces (for $G=0$) the higher Courant
bracket on $E_2=TM\oplus\mbox{\footnotesize$\bigwedge$}^2\,T^*M$ from
\eqref{courant}. Finally \eqref{c5} is the Bianchi identity which simply states that the 4-form $G$ is closed,
	\be 
	\partial_{[i}G_{jklm]}=0 \ ,
	\ee  
or $\dd G=0$, implying that $G$ defines a class in the degree~$4$
de~Rham cohomology of $M$. 

	The corresponding threebrane sigma-model becomes
	\begin{align}
	S[X,\a,A,F]&=\int_{\S_4}\,\big(F_i\w(\dd X^i+A^{i})-\a_i\w\dd A^i \nn\\ &\hspace{1cm}\qquad+\sfrac 1{4!}\,G_{ijkl}(X)\,A^i\w A^j\w A^k\w A^l\big)~.
	\end{align}
	In accord with the discussion at the end of Section \ref{sec42}, the Wess-Zumino term is associated to the higher Courant bracket on $E_2=TM\oplus \mbox{\footnotesize$\bigwedge$}^2\, T^{\ast}M$ by means of the relation 
	\be 
	\sfrac 14 \,G_{ijkl}(X)=X^{\ast}\big(\langle e_i,\langle e_j,[e_k,e_l]\rangle\rangle\big)~, 
	\ee   
	where the bracket and the bilinear form are given by \eqref{agg1} and \eqref{bilinear2} respectively.
	It is useful to add a boundary term 
	\be 
	S_{\partial}[X,A]=\oint_{\partial\S_4}\, \sfrac 12\, g_{ij}(X)\,A^i\w\ast \,A^j~,
	\ee 
	where the Hodge duality operation $\ast$ is in three dimensions, sending 1-forms to 2-forms. A topological boundary term of the form 
	\be 
	S_{\partial,\text{top}}[X,A,\a]=\oint_{\partial\S_4}\, h^i_j(X)\, \a_i\w A^{j}
	\ee 
	is also possible, as well as boundary terms of type $\a\w\ast\,\a$ and $A\w A\w A$, however we do not include them in this example.
	The field equation for the auxiliary 3-form $F_i$ gives
	\be 
	A^i=-\dd X^i~,
	\ee 
	which with $G=\dd C$ leads locally to the M2-brane action
	\be 
	S_{\partial}[X]=\oint_{\partial\S_4} \, \big(\sfrac 12\, g_{ij}\,\dd X^i\w\ast\,\dd X^j+\sfrac 1{3!}\,C_{ijk}\,\dd X^i\w\dd X^j\w\dd X^k\big)~.
	\ee  
	This is recognized as the action for a closed M2-brane coupled to a 3-form $C$-field, whose field strength is the 4-form $G$-flux. This example, without the metric term, was considered in 
	\cite{Ikeda:2010vz}, and recently in more detail in \cite{Kokenyesi:2018ynq} where the double dimensional reduction of the Wess-Zumino term for wrapped threebranes along the M-theory circle is shown to reproduce the standard Wess-Zumino membrane coupling to an NS--NS $H$-flux.
	
\paragraph{M-theory on twisted tori.}

Motivated by Scherk-Schwarz reductions of M-theory on twisted tori,
studied in detail in \cite{Hull:2006tp}, we can choose $\rho^i{}_j$ to
be equal to the components of a globally defined coframe
${\sf E}^{i}{}_{j}(X)$ for a twisted torus.{\footnote{As explained in
    e.g.~\cite{Chatzistavrakidis:2018ztm}, one has to choose an
    isomorphism ${\sf E}:TN\to TM$ between the tangent bundles of $M$ and the twisted
    torus $N$ with corresponding components ${\sf E}^{m}{}_{i}$, whose inverse ${\sf E}^{-1}:TM\to TN$ has
    components ${\sf E}_m{}^i$, and identify the anchor with those. We
    present a short-cut version of this construction here, hoping that
    no confusion is caused.}} The simplest example, but by no means
the only one, is to take a twisted 4-torus which is a trivial circle
bundle over the three-dimensional Heisenberg nilmanifold, see for example \cite{Blair:2014zba,Bosque:2016fpi,Gunaydin:2016axc,Kupriyanov:2017oob,Lust:2017bgx,Lust:2017bwq}. Then $T^{i}{}_{jk}$ is simply given by \eqref{geometricflux} and it is constant, corresponding to 
	the structure constants of the associated nilpotent Lie algebra. Due to the Jacobi identity, the condition \eqref{c4} is 
	identically satisfied. One may further choose $G=0$, in which case \eqref{c5} is also an identity and all 
	conditions are solved. 

The sigma-model is 
	\begin{align}
	S[X,\a,A,F]=\int_{\S_4}\,\big(F_i\w(\dd
          X^i+{\sf E}^{i}{}_{j}\,A^{j})-\a_i\w\dd A^i+ \sfrac 1{2}\,T^{i}{}_{jk}\,\a_i\w A^j\w A^k\big)~,
	\end{align}
	where the last term corresponds to 
	\be 
	\sfrac 12 \, T^{i}{}_{jk}=X^{\ast}\big(\langle e^{i},[e_j,e_k]\rangle\big)~,
	\ee 
	and we add the same boundary term as before. The equation of
        motion for $F_i$ yields 
	\be 
	A^{i}=-{\sf E}^{i}=-{\sf E}_j{}^i\,\dd X^j \qquad \text{with}\quad \dd {\sf E}^{i}=-\sfrac 12\, T^{i}{}_{jk}\, {\sf E}^{j}\w {\sf E}^{k}~,
	\ee 
where we used the Maurer-Cartan equations.
	Inserting this in the action, the bulk terms cancel completely and the boundary term becomes 
	\be 
	S_{\partial}[X]=\oint_{\partial\S_4}\, \sfrac 12\, g_{ij}\, {\sf E}^{i}\w\ast\, {\sf E}^{j}~, 
	\ee 
	which is the correct M2-brane action for an M-theory background with purely geometric torsion flux $T^{i}{}_{jk}$. 
	
One may already consider a more general situation, by allowing a non-vanishing constant 4-form $G$. Then one would 
	simply obtain the same model decorated with an additional $G$-flux, subject to the algebraic identity 
	\be 
	T^{n}{}_{[ij}\,G_{klm]n}=0~, 
	\ee    
	due to \eqref{c5}. This is a special case of the algebraic
        Bianchi identity
        \eqref{eq:GF0}, and it is welcoming to see that the very same condition was considered in the context of Scherk-Schwarz reductions with both fluxes turned on, see \cite[Eq. (2.5)]{Hull:2006tp}, where it ensures that the flux ${\sf G}=\frac1{4!}\, G_{ijkl}\, {\sf E}^i\w{\sf E}^j\w{\sf E}^k\w{\sf E}^l$ is closed. 
	In the present context this identity is imposed by the gauge invariance of the threebrane sigma-model via the classical master equation, and it ensures closure of the Wess-Zumino term $X^*({\sf G})$. 

\paragraph{Homotopy cotangent algebroids.}

An option that has not been explored so far is the
choice of vector bundle $E=T^{\ast}M$. As we shall see, this is the analogue of the $R$-flux with Poisson structure in the case of (contravariant) Courant algebroids \cite{Asakawa:2014kua}, which leads to a 
\emph{geometric} $R$-flux. The difference is that in the Courant algebroid, the analogues of the fields 
$\a$ and $A$ are treated symmetrically because they are both degree~$1$ fields. Here this ``symmetry'' breaks down. 

We choose a local coframe $\{e^{i}\}$ for the cotangent bundle, with
dual local frame $\{e_i\}$ of the tangent bundle. According to the explanations of Section~\ref{sec42}, the worldvolume
1-form $A=A_i\,e^i$ with $A_i=A_{i\,\a}\,\dd\s^{\a}$ is now valued in
$X^{\ast}T^{\ast}M$, while the worldvolume 2-form $\a=\alpha^i\,e_i$
with $\a^{i}=\sfrac 12\, \a^{i}_{\a\b}\,\dd\s^{\a}\w\dd\s^{\b}$ is valued in $X^{\ast}TM$. In the case of exact Courant algebroids, where $E=E_1=TM\oplus T^{\ast}M$, taking instead the dual $E_1^*=T^{\ast}M\oplus TM$ does not lead to any difference but a renaming of the fields; however, this is not the case here, since the fields have different degrees. 

Let us proceed with the analysis of the conditions for a Lie algebroid up to homotopy. First, the anchor components $\rho^{i}{}_{I}$ now become $\rho^{ij}$ and they correspond to a bivector. The condition \eqref{c1} becomes 
\be 
\rho^{ij}\,S_{jk}=0~,
\ee 
which implies $S_{ij}=0$ if as before we assume that the anchor is a non-degenerate map. Once more, the condition \eqref{c2} is then automatically satisfied. On the other hand, the relation \eqref{c3} becomes
\be 
\rho^{li}\,\partial_l\rho^{jk}-\rho^{lk}\,\partial_l\rho^{ji}-\rho^{jl}\,T_l{}^{ik}=0~.
\ee 
This is solved by identifying $\rho^{ij}={\mit\Pi}^{ij}$ with the components
of a Poisson bivector ${\mit\Pi}$, satisfying $[{\mit\Pi},{\mit\Pi}]_{\rm SN}=0$ with respect to the Schouten-Nijenhuis bracket on multivector fields, and 
\be 
T_i{}^{jk}=-\,Q_i{}^{jk}:=-\,\partial_i{\mit\Pi}^{jk}~.
\ee 
This $Q$-flux satisfies the Bianchi identity \eqref{c4}, 
\be 
{\mit\Pi}^{l[i}\,\partial_lQ_j{}^{km]}=Q_j{}^{l[k}\,Q_l{}^{mi]}~.
\ee 
If we allow a non-vanishing generalized 4-form $G$, which in the present case is a tetravector, it has to satisfy the Bianchi identity \eqref{c5} which reads as
\be 
{\mit\Pi}^{l[i}\,\partial_lG^{jkmn]}+Q_l{}^{[ij}\,G^{kmn]l}=0~,
\ee 
which may be written in the suggestive form
\be 
\dd_{{\mit\Pi}}G:=[{\mit\Pi},G]_{\rm SN}=0~.
\ee 
This is reminiscent of the Bianchi identity $[{\mit\Pi},R]_{\rm SN}=0$ in
the case of the Poisson Courant algebroid. In this case the higher
Courant bracket \eqref{eq:derivedbracket} on
$E_2^*=T^*M\oplus\mbox{\footnotesize$\bigwedge$}^2\,TM$ is written using the  Lichnerowicz
differential $\dd_E=\dd_{\mit\Pi}=[{\mit\Pi},\,\cdot\,]_{\rm SN}$ defined on multivector
fields, together with the Koszul-Schouten bracket on $E=T^*M$ which for 1-forms $\tau_1,\tau_2\in\Gamma(T^*M)$ reads as
\be
[\tau_1,\tau_2]_{\mit\Pi} = {\cal L}_{\iota_{\tau_1}{\mit\Pi}}\,\tau_2 - {\cal L}_{\iota_{\tau_2}{\mit\Pi}}\,\tau_1-\dd\,\iota_{\tau_1}\iota_{\tau_2}{\mit\Pi} \ .
\ee
Then the twisting 4-vector $G$ defines a class in the degree~4
Poisson cohomology of $M$.

Now that we have satisfied all the conditions for a Lie algebroid up to
homotopy, we are all set to write down the threebrane sigma-model for
the above data. It is given by the action
\begin{align}
S[X,\a,A,F]&=\int_{\S_4}\,\big(F_i\w\dd X^i-\a^{i}\w\dd
             A_i+{\mit\Pi}^{ij}\,F_i\w A_j \nn\\
&\hspace{1cm}\qquad -\sfrac 12\, Q_{i}^{\ jk}\,\a^{i}\w A_j\w A_k
+\sfrac 1{4!}\,G^{ijkl}\,A_i\w A_j\w A_k\w A_l\big)~.
\end{align}
The equation of motion for $F_i$ gives 
\be 
\dd X^i=-{\mit\Pi}^{ij}\, A_j~,
\ee 
which can be inverted due to the non-degeneracy of the Poisson
bivector to get
\be 
A_i=-{\mit\Pi}^{-1}_{ij}\,\dd X^j~.
\ee 
Choosing local Darboux coordinates in which both $G^{ijkl}$ and ${\mit\Pi}^{ij}$ are constant, and adding a suitable boundary term, one obtains the M2-brane sigma model
\begin{align}
S_{\partial}[X]&=\oint_{\partial\S_4}\, \sfrac 12\,
                 \big(g_{ij}-{\mit\Pi}^{-1}_{ik}\,g^{kl}\,{\mit\Pi}^{-1}_{lj}\big)\,\dd
                 X^{i}\w\ast\,\dd X^{j}\nn\\ &\hspace{1cm}\qquad +\oint_{\partial\S_4}\, \sfrac 1{4!}\, G^{pqrs}\,{\mit\Pi}^{-1}_{ip}\,{\mit\Pi}^{-1}_{jq}\,{\mit\Pi}^{-1}_{kr}\,{\mit\Pi}^{-1}_{ls}\,X^{i}\, \dd X^{j}\w \dd X^{k}\w \dd X^{l}~.
\end{align}
This is a 
non-trivial example of a Lie algebroid up to homotopy that generalizes to open threebranes the Poisson $R$-flux model for the open membrane. In the present case, the 
4-form flux is controlled by a 4-vector $G$ that satisfies $[{\mit\Pi},
G]_{\rm SN}=0$. 

We stress that this is not the analogue of the \emph{non-geometric}
$R$-flux in M-theory, as it does not correspond to the lift of the
nonassociative closed string $R$-flux deformation along the M-theory circle. It is simply the analogue of the Poisson Courant algebroid \cite{Chatzistavrakidis:2015vka,Asakawa:2014kua,Chatzistavrakidis:2018ztm} at one level higher in the geometric staircase, and it is a geometric model. The analogue of the non-geometric $R$-flux will be discussed below. 

\section{Threebrane Sigma-Models and M-Theory Fluxes} 
\label{sec5}

We would now like to address the question posed in Section \ref{sec41}, and relate the general M-theory fluxes and their Bianchi identities computed in Section \ref{sec3} to the threebrane sigma-models presented in Section \ref{sec42}. This question can now be stated in more precise terms as follows: 
do the conditions \eqref{c1}--\eqref{c5} generate all the fluxes and
Bianchi identities? We will answer this question in the affirmative and thus enable ourselves to write down the corresponding threebrane sigma-model. 

The answer is that in the context of the threebrane sigma-model we
should consider the vector bundle $E=E_2=TM\oplus
\mbox{\footnotesize$\bigwedge$}^2\,T^{\ast}M$. Although this choice seems very
reasonable in view of our previous discussion, one should appreciate
that it is not the most natural choice, which is perhaps the reason
that it has not been considered before. This may be explained by
invoking the analogy with the Courant sigma-model. In that case, one
has two different worldvolume 1-forms, say $q^{i}$ and $p_i$, taking
values in dual bundles, say $L$ and $L^{\ast}$ respectively. However,
since they are of the same degree, they can be combined in a single
1-form $A^{I}$ taking values in $E_1=L\oplus L^{\ast}$, so that $L$ is a maximally isotropic subbundle of $E_1$ with respect to the symmetric contraction pairing. Then the
natural choice would be $L=TM$, which gives rise to the generalized
tangent bundle. (The second choice $L=T^{\ast}M$ leads to the same
bundle, as we already mentioned before.) On the contrary, in the
present case, the two fields taking values in dual bundles $E$ and
$E^{\ast}$ are of different degree and they cannot be combined. The
most natural choice for $E$, the direct analogue of $L=TM$ above, would
be either the tangent or cotangent bundle, the two choices being now
inequivalent. But this is exactly what we have already done in Section
\ref{sec43}. There we saw that this can account for the $G$-flux and
for the geometric torsion flux $f^{i}{}_{jk}$, but not for the rest of
the $SL(5)$ fluxes. We also saw that there is a consistent case with a
4-vector flux, which however is not one of the $SL(5)$ fluxes. 

Here our sole purpose is to find under which conditions the general
theory yields instead the full set of $SL(5)$ fluxes and nothing more.
In contrast to the case of Courant algebroids, where the
$O(d,d)$-structure is intrinsic, the structure group for a general Lie
algebroid up to homotopy is not naturally tailored for the U-duality
group $SL(5)$ in four dimensions. Thus additional projections to $SL(5)$ tensors are
needed to make contact with the $SL(5)$ fluxes. This is actually a
strength of the present formalism, as there is some room for the same
expressions below to also give a subset of the fluxes in higher
dimensions for other exceptional U-duality groups.

As in the examples of Section \ref{sec43}, based on $E=TM$ and
$E=T^{\ast}M$ respectively, we directly set $S^{IJ}=0$, since this
pairing does not play any further role in the identifications. Once
more, this assumption takes care of the conditions \eqref{c1} and
\eqref{c2}. As we now show, this means that the relevant $SL(5)$ fluxes are determined
by the structure constants of a suitable Lie algebroid on $E=E_2$, whose bracket is specified by these structure constants.

Considering the bundle $E_2$ over a four-dimensional target space $M$ means that its local basis index $I$ takes $10$ values, split into sets of $4$ and $6$ as before, namely a lower (upper) $I$ becomes either a lower (upper) $i$ or a set of upper (lower) $[ij]$ indices, with $i,j=1,2,3,4$. Based on this, we write the components of the anchor $\rho$ as 
\be 
(\rho^i{}_I)=(\rho^i{}_j,\rho^{ijk})~, 
\ee   
and for our purposes here we further identify 
\be 
\rho^i{}_j=\d^i{}_j \qquad \mbox{and} \qquad \rho^{ijk}=\sfrac 12\, {\mit\Omega}^{ijk}~,
\ee 
where we assume that $\rho^{ijk}$ is a completely antisymmetric
3-vector, but not necessarily a Nambu-Poisson tensor.
Then the condition \eqref{c3} yields three different equations. For both $I,J$ being $i,j$ the equation is 
\be 
T^{m}{}_{ij}+\sfrac 12\, {\mit\Omega}^{mkl}\,T_{klij}=0~, 
\ee 
which is precisely the expression \eqref{exflux2}, provided we make the identifications $T^{k}{}_{ij}=F^k{}_{ij}$ and 
$T_{ijkl}=G_{ijkl}$. The latter identification is non-trivial, since $T_{ijkl}$ is not completely antisymmetric \emph{a priori}, but instead it is a \emph{reducible} mixed-symmetry tensor of type $(2,2)$. This means that in order to make contact with the $G$-flux, we consider only the irreducible fully antisymmetric component to be non-vanishing. This is a legitimate assumption, as long as the consistency conditions are satisfied, which is the case here. In a certain sense, it corresponds to a projection to $SL(5)$ representations. 

Second, for $I=i$ and $J$ being (upper) $[jk]$ 
(or vice-versa), the equation we obtain is
\be \label{exflux3b}
T_{i}{}^{jkl}=\sfrac 12\, \partial_i {\mit\Omega}^{jkl}-\sfrac 12\, {\mit\Omega}^{lmp}\,T_{mpi}{}^{jk}~.
\ee 
This instructs us to identify $T_i{}^{jkl}=Q_i{}^{jkl}$ and $T_{mpi}{}^{jk}=\widetilde F_{mpi}{}^{jk}$. 
This looks like \eqref{exflux4}, provided that we manage to fix $\widetilde F$ properly with the remaining equations. Third, taking $I$ to be (upper) $[ij]$ and $J$ to be (upper) $[kl]$, we obtain
\be \label{exflux6b}
T^{mijkl}+\sfrac 12\, {\mit\Omega}^{mpq}\,T_{pq}{}^{ijkl}=\sfrac 14\, {\mit\Omega}^{nkl}\,\partial_n{\mit\Omega}^{mij}-\sfrac 14\, {\mit\Omega}^{nij}\,\partial_n{\mit\Omega}^{mkl}~.
\ee 
This expression indicates the identifications $T^{ijklm}=R^{jk,lm,i}$ and $T_{pq}{}^{ijkl}=\widetilde Q_{pq}{}^{ij,kl}$. Whether or not it is identical to \eqref{exflux6} remains to be shown, provided we are able to fix $\widetilde Q$ from what follows. 

We now move on to the condition \eqref{c4}, whose middle term is zero
by assumption. Taking $I,K,L$ as single indices and $J$ as a doubled index, we obtain directly the Bianchi identity \eqref{exbi1} with the same identifications as above. This is then solved by taking $\widetilde F$ to be as in \eqref{exflux3}, in which case indeed \eqref{exflux3b} becomes identical to \eqref{exflux4}. Similarly, from \eqref{c4} we also identify $\widetilde Q$ as in \eqref{exflux5}, and thus \eqref{exflux6b} is identical to \eqref{exflux6}. 

Finally, in the present context \eqref{c5} is redundant and we can
take a vanishing twist $G_{IJKL}=0$. By \eqref{GWZ}, the vanishing locus of $G$ defines an isotropic subbundle of $E=TM\oplus
\mbox{\footnotesize$\bigwedge$}^2\,T^{\ast}M$. It is precisely on this isotropic subbundle that the Bianchi identities coming from the higher Courant bracket in Section~\ref{sec32} give the second condition \eqref{c4} coming from gauge invariance of the threebrane action with our projection.

Thus we conclude that indeed the sought-for equations for the fluxes and Bianchi identities are the conditions \eqref{c1}--\eqref{c5} under the above identifications. This provides a correspondence between the higher Courant bracket and this special case of the general AKSZ threebrane sigma-model. Let us therefore use this correspondence to write down the sigma-model explicitly. 
First, the 1-form $A$ and the 2-form $\a$ have components 
\bse
\begin{align}
A^I&=(A^i,A_{ij})=:(q^i,p_{ij})~, \\[4pt]
\a_I&=(\a_i,\a^{ij})=:(p_i,q^{ij})~,
\end{align}\ese
involving 1-forms taking values in (the pullback bundles of) $TM$
and $\mbox{\footnotesize$\bigwedge$}^{2}\,T^{\ast}M$, and 2-forms taking values in $T^{\ast}M$
and $\mbox{\footnotesize$\bigwedge$}^2\,TM$ respectively. It would therefore appear
that we have introduced an overabundance of worldvolume fields. However, our assumption that only the fully antisymmetric
component of the reducible tensor $T_{ijkl}$ survives and is identical
to the $G$-flux indicates that the 2-form $q^{ij}$ is decomposable,
namely $q^{ij}=c\, q^{i}\w q^{j}$, where we choose a convenient
scaling factor
$c\in\R^\times$. What is more, the fact that we restrict the target space
dimension to be four, and that the fluxes $\widetilde F$ and
$\widetilde Q$ obey the relations \eqref{ftildef} and \eqref{qtildeq},
dictates that $q^i\w p_{ij}=\frac 1{\tilde{c}} \, p_j$, where
$\tilde c\in\R^\times$. Thus we are not dealing with an arbitrary threebrane
sigma-model of the type discussed in Section~\ref{sec4}, but rather with
a special case of it dictated by the relation to the M-theory
fluxes. In the quantum theory, this would mean projecting the domain
of the path integral to worldvolume field configurations
constrained in this way.

Therefore, putting the above ingredients together, the sought-for threebrane sigma-model reads explicitly as 
\begin{align}
S&=\int_{\S_4}\,\Big(F_i\w\dd X^i-\tilde{c}\,q^{i}\w p_{ij}\w\dd q^j-c\, q^{i}\w q^{j}\w\dd p_{ij}+F_i\w q^{i}+\sfrac 12\, {\mit\Omega}^{ijk}\,F_i\w p_{jk} \nn\\ 
& \hspace{1cm} \qquad + \big(3\,c+\mbox{$\frac {\tilde c}{2}$}\big)\,
  F^{m}{}_{jk}\, q^i\w q^j\w q^k\w p_{im}+\sfrac c{2}\,
  G_{ijkl}\,q^{i}\w q^{j}\w q^{k}\w q^{l} \nn \\
& \hspace{1cm} \qquad + \big(\tilde c+2\, c\big)\, Q_l{}^{ijk}\,q^{l}\w q^{m}\w p_{mi}\w p_{jk}\,+\sfrac c{2}\, R^{jk,lm,i}\,q^{n}\w p_{ni}\w p_{jk}\w p_{lm}\Big)~, \label{qp3special}
\end{align}
up to boundary terms. This is indeed a sigma-model that contains all
types of fluxes $G$, $F$, $Q$ and $R$.
Due to the conditions we have imposed, it depends only on two first
order fields $q^{i}$ and $p_{ij}$, which are both 1-forms on
$\S_4$. One could think of them as the first order variables for spacetime
coordinates $X^{i}$ and their dual wrapping coordinates $\widetilde{X}_{ij}$, although a truly $\widetilde{X}$-inclusive model requires extension of the base manifold $M$ as in exceptional field theory, which is not the case here. With our specific projections in the threebrane sigma-model, it is straightforward if a bit lengthy to check that the projected threebrane action \eqref{qp3special} is invariant under the corresponding restriction of the gauge transformations \eqref{gt1}--\eqref{gt4}.

We now observe that the $G$-flux model may be obtained in an alternative way using \eqref{qp3special}. 
One simply takes ${\mit\Omega}=0$, i.e. the anchor to be the projection to
the tangent bundle of $M$, and the only non-vanishing flux to be
$G$. The field equation for $F_i$ leads to $q^{i}=-\dd X^{i}$
and the threebrane sigma-model reduces on-shell to 
\be
S_{G}[X]=\oint_{\partial\S_4}\, \sfrac 12\, g_{ij}\,\dd X^{i}\w \ast\,\dd
X^{j}+\int_{\S_{4}}\, \sfrac 1{4!}\, G_{ijkl}\,\dd X^{i}\w \dd X^{j}\w \dd X^{k}\w \dd X^{l}~, \label{qp3specialG}
\ee
where we consider the case that the threebrane has a non-empty closed M2-brane boundary and, using $\tilde c \, q^{i}\w p_{ij}=p_j$ and $c=\frac 1{12}$, we imposed the boundary condition 
\be 
p_i= -\,{6\,\tilde c}\, g_{ij}\,\ast\dd X^{j} \qquad \text{on}\quad  \partial\S_4~.
\ee  
This is the same as the sigma-model with $G$-flux that we obtained in a different way in Section~\ref{sec43}. Here the symmetric term corresponds to the upper-left entry of the generalized metric \eqref{exgm} for~$C=0$.

Alternatively, one may consider the completely dual situation, where
the anchor maps to the tangent bundle only through the trivector $\mit\Omega$,
i.e. taking $\rho^{i}{}_{j}=0$. For simplicity, let us suppose that
${\mit\Omega}$ is a constant non-degenerate trivector, and that the only non-vanishing (constant) flux is $R$. Then the field equation for $F_i$ is 
\be 
\dd X^{i}=-\sfrac 12\, {\mit\Omega}^{ijk}\,p_{jk}~.
\ee 
This implies ${\mit\Omega}^{ijk}\,\dd p_{jk}=0$ for all $i$, and hence, by our
assumption that ${\mit\Omega}^{ijk}$ is non-degenerate, that $\dd
p_{jk}=0$. This means that locally we can define functions
$\widetilde{X}_{ij}$ on $\S_4$ through
\be 
p_{ij}=\dd \widetilde{X}_{ij}~.
\ee 
This is not meant to be a wrapping coordinate, however the notation is
indicative of what one would expect in the case of an extended base
manifold. Then on-shell the threebrane sigma-model reduces to 
\bea 
S_{R}[\widetilde{X}]=\oint_{\partial\S_4}\, \sfrac 12\, g^{ijkl}\,\dd \widetilde{X}_{ij}\w\ast\,\dd \widetilde{X}_{kl}+\int _{\S_4}\, \sfrac 1{4!}\,R^{jk,lm,i}\,q^{n}\,\w\dd\widetilde{X}_{ni}\w\dd\widetilde{X}_{jk}\w\dd\widetilde{X}_{lm}~,
\eea 
where, taking into account $q^{ij}=\frac 1{12}\,q^{i}\w q^{j}$, we imposed the boundary condition
\be 
q^{ij}=\sfrac{1}{12\,\tilde{c}}\, g^{ijkl}\,\ast\dd\widetilde{X}_{kl} \qquad \mbox{on} \quad \partial\Sigma_4~,
\ee 
with
\be 
g^{ijkl}=g^{ik}\,g^{jl}-g^{il}\,g^{jk}~.
\ee 
This is indeed the metric naturally appearing in M2-brane duality rotations \cite{Duff:1990hn}, and it also corresponds to the lower-right entry of the generalized metric \eqref{exgm}.

An integrability problem in realizing the full $SL(5)$ U-duality group in the worldvolume theory for a closed M2-brane was identified in~\cite{Duff:2015jka}, based on previous work of~\cite{Duff:1990hn}. It would be interesting to see whether the simple constructions we
 presented here can be generalized to include the M2-brane wrapping modes and shed some light on the problem of defining a manifestly U-duality invariant sigma-model. 
 
\section{${SL(5)}$ Exceptional Field Theory Fluxes}
\label{sec6}

Going one step further, we would now like to determine in a systematic
way the fluxes in $SL(5)$ exceptional field theory, as first discussed in
\cite{Blair:2014zba}. In that case, the base manifold $M$ is extended
to include coordinates conjugate to both the momentum and wrapping
modes of closed M2-branes. In the present case the extended space
${\cal M}$ is $10$-dimensional with local coordinates
\be 
x^{I}=(x^{i}, \tilde{x}_{ij})~.
\ee 
Explicitly, $x^I=x^{\bar a\bar b}=-x^{\bar b\bar a}$ are coordinates
in the antisymmetric representation $\boldsymbol{10}$ of
$SL(5)$ with $\bar a,\bar b=1,2,3,4,5$. The spacetime coordinates are given
by $x^{i5}=-x^{5i}=x^i$ with $i=1,2,3,4$
and the wrapping coordinates are given by dualization $\tilde
x_{ij}=\frac12\,\epsilon_{ijkl}\, x^{kl}$ in four dimensions. A field in the
antisymmetric representation of $SL(5)$ is a section of the generalized tangent bundle
$E_2=TM\oplus\mbox{\footnotesize$\bigwedge$}^2\,T^*M$, so a local model
for the extended space may be taken to be the total space ${\cal M}=\mbox{\footnotesize$\bigwedge$}^2\,T^*M$ of the
bundle of 2-forms on $M$, with $x^i$ local coordinates on the base $M$
and $\tilde x_{ij}=-\tilde x_{ji}$ local fibre coordinates. 
Correspondingly, dual derivatives are defined in addition to the standard ones,
\be 
\partial_I=(\partial_i,\tilde{\partial}^{ij}) \ .
\ee  

It is instructive at this point to recall that the local symmetries of exceptional field theory are generated by a generalized Lie derivative 
\be 
{\mathscr L}_{\xi}A^{I}=\xi^{J}\,\partial_{J}A^{I}-A^{J}\,\partial_{J}\xi^{I}+Y^{IJ}_{KL}\,A^{K}\,\partial_{J}\xi^{L}~,
\ee 
where $\xi^{I}$ is a gauge generator for generalized diffeomorphisms
and $Y^{IJ}_{KL}$ is an invariant tensor of the U-duality group, here $SL(5)$. It reads as 
\be 
Y^{IJ}_{KL}=\epsilon^{\bar aIJ}\,\epsilon_{\bar aKL}~.
\ee 
The role of this tensor becomes evident when closure of the algebra of gauge transformations is imposed. This happens when a section condition is satisfied, which reads as 
\be \label{eq:sectioncondition}
Y^{IJ}_{KL}\,\partial_I\otimes\partial_J=0~.
\ee 
This operator equation expresses the fact that when it acts on any
field or product of fields the result should vanish. Restricting to
fields satisfying this section condition, one finds
\be 
[{\mathscr L}_{\xi_1},{\mathscr L}_{\xi_2}]={\mathscr L}_{\cbral \xi_1,\xi_2\cbrar}~,
\ee 
where the bracket appearing on the right-hand side is defined as 
\be \label{higherC}
\cbral\xi_1,\xi_2\cbrar=\sfrac 12\, ({\mathscr L}_{\xi_1}\xi_2-{\mathscr L}_{\xi_2}\xi_1)~.
\ee 
This bracket is the $SL(5)$ covariantization of the higher Courant
bracket \eqref{courant}, in the same way that the $O(d,d)$
covariantization of the standard Courant bracket is the C-bracket of
double field theory \cite{Hull:2009zb}. Indeed, when solving the
section condition \eqref{eq:sectioncondition} by setting the dual derivatives $\tilde\partial^{ij}$ to zero, the bracket of
\eqref{higherC} becomes precisely the higher Courant bracket
\eqref{courant}, and moreover ${\mathscr
  L}_{\xi}\,\cdot\,=\xi\,\circ\,\cdot\,$ is given by the higher
Dorfman bracket \eqref{dorfman}.
This is exactly how this bracket was originally constructed in
\cite{Berman:2011cg} and used in \cite{Bosque:2016fpi} to relate the
geometric formulation of $SL(5)$ exceptional field theory to the
embedding tensor formalism of seven-dimensional gauged supergravity,
where the embedding tensor is related to the fluxes and in turn to the
structure constants of the algebra of generalized diffeomorphisms. 

For our purposes, we employ the procedure given in \cite{Blumenhagen:2013hva} for double field theory fluxes. A representation of the algebra \eqref{agg1}--\eqref{agg3} can be given by the Lie bracket of certain vector
fields which are sections of the tangent bundle $T{\cal M}$ of the extended manifold of
exceptional field theory.  We introduce two such fields in the
holonomic basis, given by
\bse\begin{align}\label{newa}
D_i &= \partial_i+\sfrac 12\, C_{ijk}\,\tilde\partial^{jk}~,\\[4pt]
\tilde D^{jk} &= \sfrac 12\,\tilde\partial^{jk}+\sfrac 12\,{\mit\Omega}^{jkl}\,D_l~,
\end{align}\ese
and calculate their Lie brackets. Here the components of the 3-form
$C=\frac16\, C_{ijk}\, \dd x^i\w\dd x^j\w\dd x^k$ and the 3-vector
${\mit\Omega}=\frac16\, {\mit\Omega}^{ijk}\,\partial_i\w\partial_j\w\partial_k$
can depend on both physical and wrapping coordinates. The result is 
\bse\begin{align}\label{newaa}
[D_i,D_j]&=G_{ijkl}\,\tilde D^{kl}+F_{ij}{}^m\,D_m~,\\[4pt]
{[}D_i,\tilde D^{jk}]&=\widetilde F_{ilm}{}^{jk}\,\tilde D^{lm}+Q_i{}^{jkm}\,D_m~,\\[4pt]
{[} \tilde D^{ij},\tilde D^{kl}]&=R^{ij,kl,n}\,D_n+\widetilde
                                 Q_{mn}{}^{ij,kl}\,\tilde D^{mn}~,
\end{align}\ese
where we defined the exceptional field theory fluxes
\bse\begin{align}
G_{ijkl}&=4\,\partial_{[i}C_{jkl]}+2\,C_{mn[i}\,\tilde\partial^{mn}C_{jkl]}~,\\[4pt]
F_{ij}{}^m&=-\sfrac 12\,G_{ijkl}\,{\mit\Omega}^{klm}+  \tilde{\partial}^{mk}C_{ijk}~,\\[4pt]
\widetilde F_{ilm}{}^{jk}&=\sfrac 12\,G_{ilmn}\,{\mit\Omega}^{njk}-\sfrac 12\,\tilde\partial^{jk}C_{ilm}~,\\[4pt]
Q_i{}^{jkm}&=\sfrac 12\,\big(\partial_i{\mit\Omega}^{jkm}+\sfrac12\,C_{iln}\,\tilde\partial^{ln}{\mit\Omega}^{jkm}+\sfrac 12\,  {\mit\Omega}^{lnm}\,\tilde\partial^{jk}C_{iln}+{\mit\Omega}^{ljk}\,\tilde\partial^{mn}C_{iln}\nn\\
&\quad
-\sfrac 12\,{\mit\Omega}^{jkn}\,G_{inps}\,{\mit\Omega}^{psm}\big)~,\\[4pt]
\widetilde Q_{mn}{}^{ij,kl}&=\sfrac 14\,\big({\mit\Omega}^{ijp}\,G_{pp'mn}\,{\mit\Omega}^{p'kl}+{\mit\Omega}^{klr}\,\tilde\partial^{ij}C_{rmn}-{\mit\Omega}^{ijr}\,\tilde\partial^{kl}C_{rmn}\big)\nn\\
&\quad
-\sfrac 14\,\big(\delta^l_{[m}\,\partial_{n]}{\mit\Omega}^{ijk}- \delta^k_{[m}\,\partial_{n]}{\mit\Omega}^{ijl}- \delta^j_{[m}\,\partial_{n]}{\mit\Omega}^{ikl}+ \delta^i_{[m}\,\partial_{n]}{\mit\Omega}^{jkl}\big)\\
&\quad -\sfrac 18\,\big(\delta^l_{[m}\,C_{n]sp}\,\tilde\partial^{sp}{\mit\Omega}^{ijk}- \delta^k_{[m}\,C_{n]sp}\,\tilde\partial^{sp}{\mit\Omega}^{ijl}- \delta^j_{[m}\,C_{n]sp}\,\tilde\partial^{sp}{\mit\Omega}^{ikl}+ \delta^i_{[m}\,C_{n]sp}\,\tilde\partial^{sp}{\mit\Omega}^{jkl}\big)~, \nn \\[4pt]
R^{ij,kl,n}&=\sfrac 12\, \hat\partial^{i[j}{\mit\Omega}^{kln]}-\sfrac 12\, \hat\partial^{j[i}{\mit\Omega}^{kln]}-\sfrac 12\,\hat\partial^{k[l}{\mit\Omega}^{ijn]}+\sfrac 12\,\hat\partial^{l[k}{\mit\Omega}^{ijn]}-\sfrac 18\,{\mit\Omega}^{ijm}\,{\mit\Omega}^{klp}\,{\mit\Omega}^{rsn}\,G_{mprs}\nn\\
&\quad +\sfrac 14\, C_{mpr}\,\big({\mit\Omega}^{mi[j}\,\tilde\partial^{\underline{p}\underline{r}}{\mit\Omega}^{kln]}-  {\mit\Omega}^{mj[i}\,\tilde\partial^{\underline{p}\underline{r}}{\mit\Omega}^{kln]}- {\mit\Omega}^{mk[l}\,\tilde\partial^{\underline{p}\underline{r}}{\mit\Omega}^{ijn]}+ {\mit\Omega}^{ml[k}\,\tilde\partial^{\underline{p}\underline{r}}{\mit\Omega}^{ijn]}\big)\nn\\  
&\quad
+\sfrac 18\,\big({\mit\Omega}^{stn}\,{\mit\Omega}^{{i}j{r}}\,\tilde\partial^{kl}-{\mit\Omega}^{stn}\,{\mit\Omega}^{{k}lr}\,\tilde\partial^{ij}+2{\mit\Omega}^{ijt}\,{\mit\Omega}^{klr}\,\tilde\partial^{ns}\big)C_{rst}~,
\end{align}\ese
and
$ \hat\partial^{ij}=\tilde\partial^{ij}+{\mit\Omega}^{ijk}\,\partial_k$.
These expressions rely on the section condition, as also happens in
the case of double field theory.
As in the case of exceptional generalized geometry, these expressions are valid for any $d$, in which case the dimension of the extended space is $d+\frac{d\,(d-1)}{2}$, however they simplify for the physically relevant case of $d=4$, where  $\widetilde F$ and $\widetilde Q$  can be related to  $F$ and $Q$ respectively.  
 
To write these expressions in a non-holonomic frame, we introduce a vielbein $e_{a}=e_{a}{}^{i}\,\partial_i$ whose components $e_a{}^i$ can depend on both physical and
wrapping coordinates, together with the dual vector fields
$\tilde e^{ab}=e^{[a}{}_i\,e^{b]}{}_j\,\tilde\partial^{ij}$. 
Then the fluxes acquire additional terms and in the
four-dimensional case they read as
\bse \begin{align} 
G_{abcd}&=4\,\nabla_{[a}C_{bcd]}+2\,C_{ef[a}\,\widetilde{\nabla}^{ef}C_{bcd]}~, \\[4pt]
F_{ab}{}^{c}&= f_{ab}{}^{c}+C_{de[a}\,{\mit{\tilde\G}}^{de}{}_{b]}{}^{c}-\sfrac 12\,
              {\mit\Omega}^{dec}\,G_{abde}+
              \widetilde{\nabla}^{cd}C_{dab}\ ,\\[4pt]
Q_{a}{}^{bcd}&=\sfrac 12\, \partial_a{\mit\Omega}^{bcd}-\sfrac 32\, {\mit{\tilde\Gamma}}^{[bc}{}_{a}{}^{d]}+\sfrac 32\, {\mit\Omega}^{e[bc}\,f_{ae}{}^{d]}  +\sfrac 14\, C_{aef}\,\widetilde{\nabla}^{ef}{\mit\Omega}^{bcd}-\sfrac 34\,C_{efg}\,{\mit{\tilde\Gamma}}^{ef}{}_{a}{}^{[d}\,{\mit\Omega}^{bc]g}\nn\\ 
&\quad + \sfrac 12\, {\mit{\tilde\Gamma}}^{de}{}_{e}{}^{[c}\,\delta^{b]}_a+\sfrac 14\,{\mit\Omega}^{def}\,C_{fgh}\,{\mit{\tilde\Gamma}}^{gh}{}_{e}{}^{[c}\,\delta^{b]}_a
-\sfrac  14\, {\mit\Omega}^{def}\,f_{ef}{}^{[c}\,\delta^{b]}_a\nn\\
&\quad +\sfrac 14\, {\mit\Omega}^{efd}\,\widetilde{\nabla}^{bc}C_{aef}\,+\sfrac 12\, {\mit\Omega}^{bce}\,\widetilde{\nabla}^{df}C_{aef}
-\sfrac 14\, {\mit\Omega}^{e[bc}\,{\mit\Omega}^{d]fg}\,G_{aefg}~, \\[4pt]
R^{ab,cd,e}&=\sfrac 12\,\widehat{{\nabla}}{}^{a[b}\,{\mit\Omega}^{cde]}-\sfrac 12\,\widehat{{\nabla}}{}^{b[a}\,{\mit\Omega}^{cde]}-\sfrac 12\,\widehat{\nabla}{}^{c[d}\,{\mit\Omega}^{abe]}+\sfrac 12\,\widehat{\nabla}{}^{d[c}\,{\mit\Omega}^{abe]} \nn \\
&\quad+\sfrac 14\, C_{fgh}\,\big({\mit\Omega}^{fa[b}\,\widetilde{\nabla}^{\underline{g}\underline{h}}{\mit\Omega}^{cde]}-{\mit\Omega}^{fb[a}\,\widetilde{\nabla}^{\underline{g}\underline{h}}{\mit\Omega}^{cde]}-
{\mit\Omega}^{fc[d}\,\widetilde{\nabla}^{\underline{g}\underline{h}}{\mit\Omega}^{abe]}+{\mit\Omega}^{fd[c}\,\widetilde{\nabla}^{\underline{g}\underline{h}}{\mit\Omega}^{abe]}\big)\nn\\ 
&\quad +\sfrac
  18\,\big({\mit\Omega}^{fge}\,{\mit\Omega}^{{a}bh}\,\widetilde{\nabla}^{cd}
  -{\mit\Omega}^{fge}\,{\mit\Omega}^{{c}d{h}}\,\widetilde{\nabla}^{ab}+2{\mit\Omega}^{abg}\,{\mit\Omega}^{cdh}\,\widetilde{\nabla}^{ef}\big)C_{hfg}~,
\end{align}\ese  
where we defined the dual connection
\be 
{\mit{\tilde\Gamma}}^{ab}{}_{c}{}^{d}=e^{d}{}_{k}\,e^{[a}{}_i\,e^{b]}{}_j\,\tilde\partial^{ij}e^{k}{}_{c}~,
\ee 
and 
\bse\begin{align}
\widetilde{\nabla}^{ab}C_{cde}&=\tilde{\partial}^{ab}C_{cde}-{\mit{\tilde\Gamma}}^{ab}{}_{c}{}^{f}\,C_{fde}
-{\mit{\tilde\Gamma}}^{ab}{}_{d}{}^{f}\,C_{cfe}-{\mit{\tilde\Gamma}}^{ab}{}_{e}{}^{f}\,C_{cdf}~,
\\[4pt]
\widehat{\nabla}^{ab}&=\widetilde{\nabla}^{ab}+{\mit\Omega}^{abc}\,\nabla_c~,
\end{align}\ese
are the dual covariant derivatives.

Finally, let us compare our results with the $SL(5)$ fluxes described in
\cite{Blair:2014zba,Lust:2017bwq}, where a group theoretical
derivation in terms of $SL(5)$ representation theory was used. Based
on the embedding tensor formalism for gaugings of seven-dimensional
maximal supergravity~\cite{Samtleben:2005bp}, the relevant
representations of $SL(5)$ are 
$\mathbf{\overline{15}}\oplus\mathbf{\overline{40}}\oplus\mathbf{\overline{10}}$,
and therefore the fluxes should exhaust these representations. This
may be confirmed by using their branching decompositions under the embedding
$SL(5)\supset SL(4)\times\R^+$, which read as{\footnote{We refrain from presenting the additional $\R^{+}$ charges here.}}  
\bse \begin{align}
\mathbf{\overline{10}}\,\big|_{SL(4)}&=\mathbf{\overline{4}}\oplus \mathbf{6}~,\\[4pt]
\mathbf{\overline{15}}\,\big|_{SL(4)}&=\mathbf{\overline{10}}\oplus \mathbf{\overline{4}}\oplus \mathbf{1}~, \\[4pt]
\mathbf{\overline{40}}\,\big|_{SL(4)}&=\mathbf{\overline{20}}\oplus\mathbf{10}\oplus \mathbf{6}\oplus \mathbf{\overline{4}}~.
\end{align}\ese 
The available fluxes that should be matched with these representations
are the 4-form $G$-flux $G_{abcd}$, the geometric torsion flux
$f_{ab}{}^{c}$, the $Q$-flux $Q_{a}{}^{bcd}$ and the $\cal R$-flux
${\cal R}^{a,bcde}$. Clearly, the only singlet in these decompositions corresponds to the 4-form $G$-flux. 
Moreover, the $\cal R$-flux, as a mixed-symmetry $(1,4)$ tensor, lives
in one of the three $\mathbf{\overline{4}}$ representations of
$SL(4)$---the one in the decomposition of $\mathbf{\overline{40}}$. The
torsion flux $f_{ab}{}^{c}$ has $24$ components and lives in the
representations $\mathbf{\overline{20}}\oplus\mathbf{\overline{4}}$,
corresponding to its trace and traceless parts. The flux
$Q_{a}{}^{bcd}$,  antisymmetric in its upper three indices, has $16$ components, corresponding to the representations
$\mathbf{\overline{10}}\oplus\mathbf{6}$, again containing a trace and
a traceless part. What remains is the trace part of the dual torsion
flux ${\mit{\tilde\G}}^{ab}{}_{b}{}^{c}$, which contains a symmetric and an
antisymmetric part with a total of $16$ components living in the
representations $\mathbf{10}\oplus\mathbf{6}$. The last
$\mathbf{\overline{4}}$ representation corresponds to the determinant
of the seven-dimensional metric, and this exhausts all representations
of $SL(4)$ appearing above. 
Therefore, the  geometric approach based on the higher Courant bracket reproduces all the fluxes obtained using the  group theoretical approach.

\section{Conclusions and Outlook}
\label{sec7}     

Motivated by the relation between T-duality and non-geometric fluxes
in closed string theory, the properties of the Courant bracket in
generalized geometry, and the gauge structure of AKSZ-type topological
membrane sigma-models, in this paper we have investigated whether such
relations extend to the case of U-duality and fluxes in M-theory, the
higher Courant bracket in exceptional generalized geometry, and
AKSZ-type topological threebrane sigma-models. We established that
upon a certain projection to $SL(5)$ tensors, the local coordinate
form of the axioms for a specific Lie algebroid up to homotopy based
on the extended bundle $TM\oplus
\mbox{\footnotesize$\bigwedge$}^{2}\,T^{\ast}M$ coincides with the
general expressions for geometric and non-geometric fluxes in
$(7+4)$-dimensional compactifications of M-theory, together with their
Bianchi identities. The same expressions are also interpreted as the
conditions for gauge invariance and closure of gauge transformations
for a topological threebrane sigma-model, where the fluxes appear as
generalized Wess-Zumino terms. It would be interesting to understand
better the geometric features of the algebroid structure defined by our $SL(5)$ projection of
the Lie algebroid up to homotopy on $TM\oplus
\mbox{\footnotesize$\bigwedge$}^{2}\,T^{\ast}M$. Given that the higher
Courant algebroid structure on this extended bundle has a well-known
realization as an $L_\infty$-algebra, see for example~\cite{Zambon,Bi,Baraglia:2011dg}, it is an
interesting problem to see how our specific Lie algebroid up to
homotopy fits into recent discussions of the $L_\infty$-algebra
structure of gauge symmetries underlying the tensor hierarchy in exceptional field theory, see for
instance~\cite{Cederwall:2018aab,Arvanitakis,Cagnacci:2018buk}.

This observation can serve as a first step toward a new geometric understanding of the worldvolume approach to M-theory. This requires extension of the analysis presented here in several directions. 
In this paper we focused on the case of the exceptional group $SL(5)$,
which is the (continuous) U-duality group for M-theory compactified on a
four-dimensional torus. Obviously, a more complete treatment would
require studying the fluxes for lower (higher) number of external
(internal) dimensions, where the U-duality group is larger.  A
standard complication in that case is that M5-brane charges emerge,
and the extended bundles are larger and not of the type $TM\oplus
\mbox{\footnotesize$\bigwedge$}^p\,T^{\ast}M$, while the corresponding
topological sigma-model that encodes the fluxes as Wess-Zumino terms
is expected to be more complicated as U-duality now exchanges charged
objects of different dimensionalities. Some discussions of these issues
can be found in~\cite{Kokenyesi:2018ynq,Arvanitakis,Berman:2018okd}.

Another open problem is to construct a threebrane sigma-model with an
extended base space, similarly to double field theory where the
coordinates are doubled, whose generalized Wess-Zumino terms would accomodate the exceptional field theory fluxes, derived for the U-duality group $SL(5)$ in Section~\ref{sec6}. In the $SL(5)$ case this would require a
ten-dimensional extended space, and even higher-dimensional for larger
U-duality groups.  Target space exceptional field theories with
manifest U-duality invariance were constructed in recent years
in~\cite{eft0,eft1,eft2,eft3,eft4}, however the corresponding
worldvolume problem remains open \cite{Duff:2015jka}. Applying the
strategy developed in \cite{Chatzistavrakidis:2018ztm} for the
construction of T-duality invariant membrane sigma-models could be
helpful in that case, although its precise implementation is not
straightforward. In addition, one should then deal with the section
condition and understand its geometric origin in the context of
weaker algebroid structures, similar to the (pre-)DFT algebroids
defined in~\cite{Chatzistavrakidis:2018ztm} and formulated in~\cite{KokenyesiDFT}
in terms of an AKSZ-type construction. 

Finally, another issue that we did not discuss in the present paper is
the nonassociativity of the M2-brane phase space, which is deformed by
the presence of the locally non-geometric M-theory $R$-flux~\cite{Gunaydin:2016axc,Kupriyanov:2017oob}. This problem
requires the extension of the base manifold $M$ and falls in the same
line of discussion as previously. In other words, one would have to
include the wrapping coordinates of closed M2-branes in order to see
how nonassociativity manifests itself in the worldvolume
approach. Similarly, it would be interesting to find a geometric
interpretation for the fact that the phase space of M-theory, when
$R$-flux is turned on, is dimensionally reduced from eight to seven
dimensions due to the absence of momentum modes along the M-theory circle~\cite{Gunaydin:2016axc,Kupriyanov:2017oob,Lust:2017bgx,Lust:2017bwq}. 

\paragraph{Acknowledgements.} We thank
Ralph Blumenhagen, Mark Bugden, Fech Scen Khoo, Zolt\'an
K\"ok\'enyesi, Emanuel Malek, Erik Plauschinn, Felix Rudolph and  Marc Syv\"ari for
helpful discussions. We acknowledge support by COST (European Cooperation in Science and Technology) in the framework of the Action MP1405 QSPACE. The work of A.Ch. is supported by the Croatian Science Foundation Project ``New Geometries for Gravity and Spacetime'' (IP-2018-01-7615). A.Ch. and L.J. are partially supported by the European Union through the European Regional Development Fund -- The Competitiveness and Cohesion Operational Programme (KK.01.1.1.06) and by the H2020 CSA Twinning Project No. 692194 ``RBI-T-WINNING''. The work of D.L. was partially supported by the ERC Advanced Grant ``Strings and Gravity'' (Grant No.~320045). The work of R.J.S. was
supported by the Consolidated Grant ST/P000363/1 
from the UK Science and Technology Facilities Council.

\end{document}